\documentclass[showpacs,amsmath,amssymb,aps,twocolumn,longbibliography,pra]{revtex4-1}
\usepackage{graphicx}
\usepackage[colorlinks=true,citecolor=blue,linkcolor=magenta]{hyperref}
\usepackage[usenames]{color}
\usepackage{relsize}
\usepackage[english]{babel}
\usepackage{enumerate}

\newcommand {\grsim} {\ {\raise-.5ex\hbox{$\buildrel>\over\sim$}}\ }
\newcommand {\lessim} {\ {\raise-.5ex\hbox{$\buildrel<\over\sim$}}\ }

\newcommand{\qbf}{\mathbf{q}}
\newcommand{\qp}{\mathbf{q}^{\perp}}

\newcommand{\rp}{\mathbf{r}_{\perp}}
\newcommand{\ah}{\hat{a}}

\newcommand{\Eav}{E_{\rm{eff}}^{B}}

\newcommand{\Jeff}{J_{\mathrm{eff}}}

\begin{document}

\title{Parametric instabilities of interacting bosons  \\ in periodically-driven 1D optical lattices}

\author{K. Wintersperger$^{1,2,*}$, M. Bukov$^{3,*}$, J. N\"ager$^{1,2}$, S. Lellouch$^{4,5}$, \\ E. Demler$^{6}$, U. Schneider$^{7}$, I. Bloch$^{1,2,8}$, N. Goldman$^{4}$, M. Aidelsburger$^{1,2}$}

\affiliation{$^{1}$\,Fakult\"at f\"ur Physik, Ludwig-Maximilians-Universit\"at M\"unchen, Schellingstra{\ss}e 4, 80799 M\"unchen, Germany}
\affiliation{$^{2}$\,Munich Center for Quantum Science and Technology (MCQST), Schellingstr. 4, 80799 M\"unchen, Germany}
\affiliation{$^{3}$\,Department of Physics, University of California, Berkeley, CA 94720, USA}
\affiliation{$^{4}$\,Center for Nonlinear Phenomena and Complex Systems, Universit\'e Libre de Bruxelles, CP 231, Campus Plaine, 1050 Brussels, Belgium}
\affiliation{$^{5}$\,Laboratoire de Physique des Lasers, Atomes et Mol\'ecules, Universit\'e Lille 1 Sciences et Technologies, CNRS, 59655 Villeneuve d'Ascq Cedex, France}
\affiliation{$^{6}$\,Department of Physics, Harvard University, Cambridge, MA 02138, USA}
\affiliation{$^{7}$\,Cavendish Laboratory, University of Cambridge, J.~J.~Thomson Avenue, Cambridge CB3 0HE, UK}
\affiliation{$^{8}$\,Max-Planck-Institut f\"ur Quantenoptik, Hans-Kopfermann-Stra{\ss}e 1, 85748 Garching, Germany}
\affiliation{$^{*}$\,these authors contributed equally to this work}


\begin{abstract} 
Periodically-driven quantum systems are currently explored in view of realizing novel many-body phases of matter. This approach is particularly promising in gases of ultracold atoms, where sophisticated shaking protocols can be realized and inter-particle interactions are well controlled. The combination of interactions and time-periodic driving, however, often leads to uncontrollable heating and instabilities, potentially preventing practical applications of Floquet-engineering in large many-body quantum systems. In this work, we experimentally identify the existence of parametric instabilities in weakly-interacting Bose-Einstein condensates in strongly-driven optical lattices through momentum-resolved measurements. Parametric instabilities can trigger the destruction of weakly-interacting Bose-Einstein condensates through the rapid growth of collective excitations, in particular in systems with weak harmonic confinement transverse to the lattice axis.
\end{abstract}

\maketitle

Floquet engineering has proven to be a powerful technique for the design of novel quantum systems with tailored properties, unattainable in conventional static systems~\cite{Goldman:2014bva,Bukov:2015gu,Eckardt:2017hca}. It is based on the design of time-periodic systems, whose stroboscopic evolution is governed by an effective time-independent Hamiltonian featuring the desired properties. Floquet engineering is captivating due to its conceptual simplicity and its potentially far-reaching applications for engineering novel states of matter. For instance, it has been used to manipulate the electronic properties of solid-state systems~\cite{Fausti:2011dy,Wang:2013fe,Mahmood:2016bu,Cavalleri:2017fg}, to realize time crystals~\cite{Zhang:2017ci,Choi:2017ho}, to engineer artificial magnetic fields and topological Bloch bands in cold atoms~\cite{Struck:2011is,Aidelsburger:2011hl,Miyake:2013jw,Jotzu:2014kz,Aidelsburger:2015hm,Cooper:2018tr}, photonics~\cite{Rechtsman:2013fe,Hafezi:2013jg,Ozawa:2018us} and superconducting circuits~\cite{Roushan:2016iu}, to generate density-dependent gauge fields~\cite{Clark:2018fi,Gorg:2018de} and to explore the rich physics of lattice gauge theories~\cite{Schweizer:2019us}.

\begin{figure}[!htb]
\includegraphics{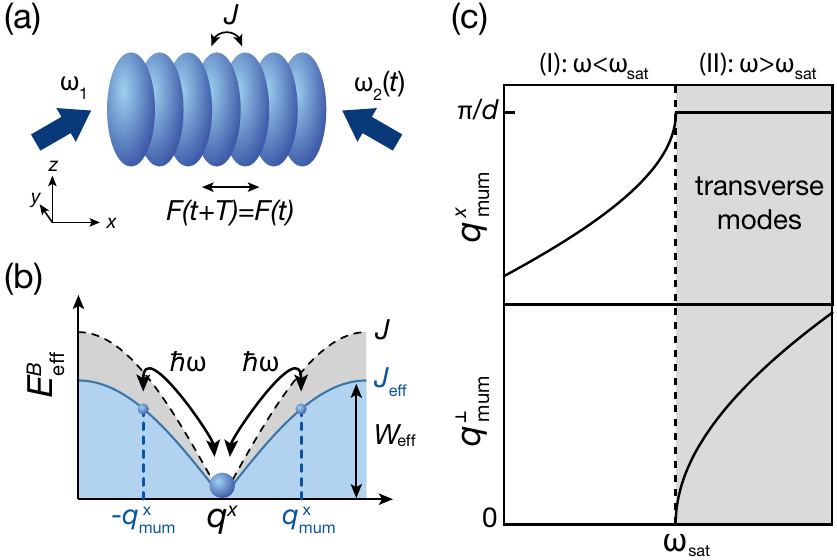}
\vspace{-0.cm} \caption{Illustrations of the experimental setup and the properties of parametric instabilities. (a) Schematic of the driven 1D optical lattice with tunneling $J$, generated by two laser beams with frequencies $\omega_{1,2}$, and weak harmonic transverse confinement. Modulating $\omega_2(t)$ periodically with frequency $\omega$ generates a force $F(t)$ with period $T=2\pi/\omega$. (b) Illustration of the effective 1D Bogoliubov dispersion $E_{\text{eff}}^B(q^x,J_{\text{eff}})$, with width $W_{\text{eff}}$ reduced according to the Floquet-renormalized tunnel coupling $J_{\text{eff}}$ (blue) compared to the static lattice case (gray). Parametric resonances at $E_{\text{eff}}^B=\hbar\omega$ induced by the modulation lead to instabilities centered around the most unstable mode $q^x_{\text{mum}}$; $\hbar=h/(2\pi)$ is the reduced Planck's constant.
(c) Momentum of the most unstable mode $\mathbf{q}_{\text{mum}} = (q^x_{\text{mum}}, \mathbf{q}^\perp
_{\text{mum}})$ as predicted by Bogoliubov theory, which shows a clear separation between lattice (I) and transverse (II) degrees of freedom, that occurs at the saturation frequency $\hbar\omega_{\text{sat}} \approx W_{\rm{eff}}$; $d$ is the lattice constant.
}
\label{Fig_1}
\end{figure}

The complex interplay between periodic driving and interactions poses theoretical and experimental challenges. In time-periodic systems, energy conservation is relaxed due to the possibility to absorb and emit energy quanta from the drive, and any driven ergodic system is expected to eventually heat up to infinite temperatures~\cite{Lazarides:2014cl,DAlessio:2014fg}. Recent experiments~\cite{Jotzu:2015bra,Kennedy:2015dw,Reitter:2017ij,CabreraGutierrez:2018tc,Messer:2018tk} have addressed this problem for interacting atoms in shaken optical lattices. In particular, it has been shown~\cite{Kitagawa:2011fj,Choudhury:2015dm,Bilitewski:2015kx,Bilitewski:2015cs,Reitter:2017ij} that heating rates are well captured by a Floquet Fermi's Golden Rule (FFGR) approach, if they are evaluated at sufficiently long times. This approach suggests that the long-time dynamics is dominated by incoherent two-body scattering processes~\cite{Bilitewski:2015kx}. In contrast, the onset of heating in bosonic systems is expected to be triggered by coherent processes~\cite{Kramer:2005dk,Creffield:2009cp,Bukov:2015dv,Lellouch:2017it}. 

First evidence for parametric instabilities in amplitude-modulated optical lattices has been found indirectly via spectroscopic measurements~\cite{Stoferle:2004ji,Kramer:2005dk}. In this work, we directly reveal the existence and nature of parametric instabilities by measuring the momentum distribution of weakly-interacting bosons in a periodically-driven one-dimensional (1D) optical lattice (Fig.~\ref{Fig_1}a). These instabilities exist whenever the energy quantum $\hbar \omega$ associated with the drive matches the energy of a collective excitation, as dictated by the effective Bogoliubov dispersion $E^B_{\text{eff}}(\mathbf{q},J_{\text{eff}})$, where $\mathbf{q}$ denotes the momentum of the excitation and $J_{\text{eff}}$ is the effective tunnel coupling renormalized by the drive~\cite{Lignier:2007du} (Fig.~\ref{Fig_1}b). In a strictly 1D lattice, the strongest instability occurs at the two-photon resonance, $2\hbar\omega \!=\! 2E_{\text{eff}}^B(q^x_{\text{res}},J_{\text{eff}})$. In the single-band approximation, there exists a stable parameter regime without parametric instabilities for $\omega > \omega_{\text{sat}} \approx W_{\rm{eff}}/\hbar$, where $W_{\rm{eff}}=\sqrt{4|J_{\text{eff}}|(4|J_{\text{eff}}|+2g)}$ is the bandwidth of the effective Bogoliubov dispersion and $g$ is the interaction energy (Ref.~\cite{Lellouch:2017it} and App.~\ref{subsec:AnaRes}). 
In a 3D system with weak harmonic confinement transverse to the lattice axis (Fig.~\ref{Fig_1}a), however, no stable parameter regime exists~\cite{Choudhury:2015dm,Bilitewski:2015kx,Bilitewski:2015cs,Reitter:2017ij,Kramer:2005dk,Lellouch:2017it} and parametric instabilities occur via the closely-spaced transverse modes~\cite{Stoferle:2004ji,Kramer:2005dk,Lellouch:2017it} [regime (II) in Fig.~\ref{Fig_1}c]. 
Moreover, our numerical simulations indicate that a harmonic confinement along the lattice axis further prevents the existence of a stable region, even in a true 1D geometry~(App.~\ref{sec:SI_trap}).

The appearance of instabilities is common in static weakly-interacting bosonic lattice systems. For instance, Landau instabilities can occur when the condensate is prepared at a finite quasimomentum, where the effective mass in the band structure is negative; such configurations can also display dynamical instabilities, where Bogoliubov excitations grow exponentially~\cite{Wu:2001ib,Modugno2004,Fallani:2004hb,LDeSarlo:2005co,Gemelke:2005dk,Campbell:2006di,Michon:2017tw}. We emphasize that the origin of such instabilities is different compared to those revealed in this work. These instabilities exist even for a condensate initially at rest, in contrast parametric instabilities originate from the time-dependent nature of the drive. While most experiments have studied the appearance of instabilities via decoherence or loss-rate measurements. We note that instabilities are not necessarily detrimental but can result in interesting phenomena, such as parametric amplification, four-wave mixing~\cite{Gemelke:2005dk,Campbell:2006di,Michon:2017tw,Clark:2017hw} and pattern formation~\cite{Staliunas:2004ci,Engels:2007cb,Fu:2018vd,Zhang:2019}.

\begin{figure}
\includegraphics{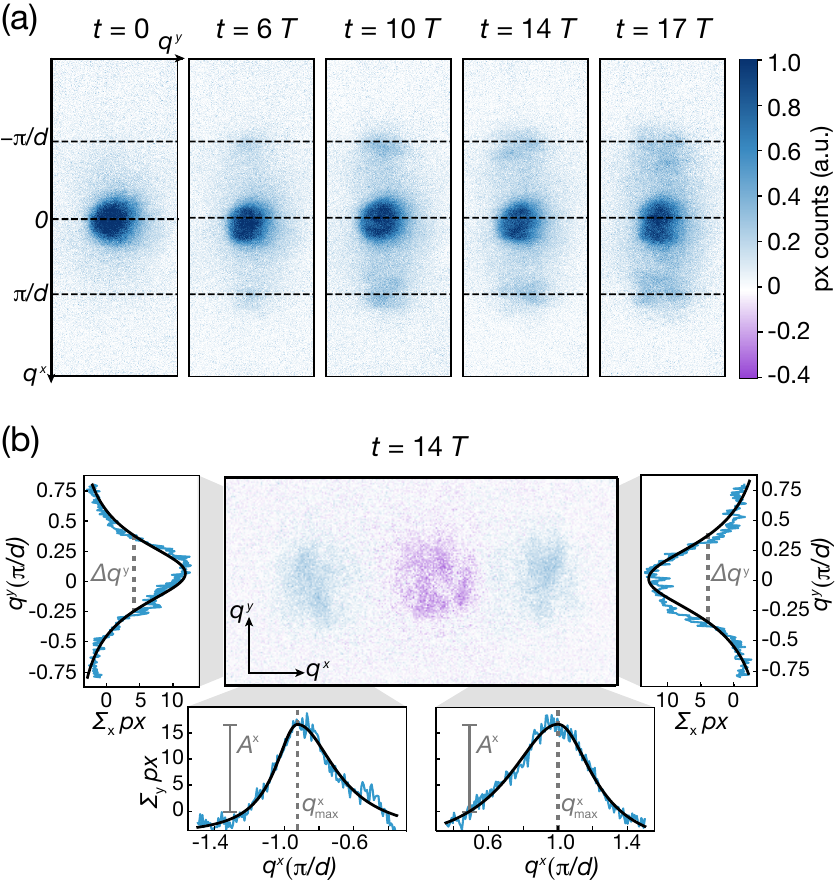}
\vspace{-0.cm} \caption{Momentum-resolved images of the most unstable mode for $\alpha=1.78$, $\omega/(2\pi) = 720 \text{Hz}$ and $\varphi=0$. (a) Time series of absorption images after $6 \text{ms}$ TOF. The edges and the center of the first Brillouin zone (BZ) are marked by dashed lines. (b) Difference image at $t=14\,T$, obtained by subtracting the condensate profile at $t=0$. The lower panels show the 1D profiles along $q^x$, resulting from summation of the difference image perpendicular to the lattice axis for the left and right excitation peak together with asymmetric Lorentzian fits (black solid lines), used to determine the amplitude $A^x$ and position $q^x_{\text{max}}$ (dashed lines) of the excitation peaks~\cite{supplements}. The left and right plot show the profiles resulting from summation along the lattice axis with Lorentzian fits to determine the transverse full-width at half maximum $\Delta q^y$ (dashed lines) of the peaks. Here, we integrate the density profile in a region of interest containing the left or right peak only.
}
\label{Fig_2}
\end{figure}

The experiment starts by loading an almost pure Bose-Einstein condensate (BEC) of about $N=3.7(4)\times10^5$ $^{39}$K atoms within $100\,$ms into a 1D optical lattice aligned along the $x$ axis, with lattice constant $d=425\,$nm and depth $V_{\mathrm{lat}}=11.0(3)\,E_{\text{R}}$, where $E_{\text{R}}=h^2/(8 m d^2)=h \times 7.1\,$kHz is the recoil energy and $m$ the mass of an atom. Additional confinement is provided by an optical dipole trap. The harmonic trapping frequencies of the combined potential are $\omega_r/(2\pi)= 26(2)\,$Hz in the $xy$-plane and $\omega_z/(2\pi)= 204(3)\,$Hz in the vertical direction. The lattice is created by interfering two laser beams with $\lambda = 736.8\,$nm under an angle of $120^{\circ}$. Its position is modulated by varying the frequency of one lattice laser beam (Fig.~\ref{Fig_1}a), $\omega_2(t)\!=\!\omega_1\!+\!2\pi \nu \sin (\omega t \!+\! \varphi)$, where $\varphi$ is the phase of the drive. The lattice modulation is turned on suddenly in order to be able to observe the presence of collective excitations after few modulation periods (we verified that ramping up the modulation amplitude within five cycles does not modify our main results). We hold the atoms in the modulated lattice for integer multiples of the driving period and determine the momentum distribution by performing bandmapping (the modulation is turned off abruptly followed by a $100 \mu \text{s}$-long linear ramp-down of the lattice) and subsequent time-of-flight (TOF) imaging (Fig.~\ref{Fig_2}a).

In the reference frame of the lattice, the modulation leads to a time-varying force $F(t) = F_0 \cos (\omega t + \varphi)$, with $F_0\!=\!m d \nu \omega$ \cite{Lignier:2007du} and the time-dependent tight-binding Hamiltonian describing the dynamics takes the form:
\begin{align}
\label{eq:hamiltonian}
\hat{H}(t) = & \int_{\mathbf{r}_\perp} \left[-J \sum_{\langle ij\rangle} \left( \hat{a}^{\dagger}_{i,\mathbf{r}^\perp} \hat{a}_{j,\mathbf{r}^\perp} + \hat{a}^{\dagger}_{j,\mathbf{r}^\perp} \hat{a}_{i,\mathbf{r}^\perp} \right)\right. \\
& + \frac{U}{2} \sum_j \hat{n}_{j,\mathbf{r}^\perp} \left(\hat{n}_{j,\mathbf{r}^\perp} -1\right) \nonumber\\
& +\left. K \cos (\omega t + \varphi) \sum_j j \, \hat{n}_{j,\mathbf{r}^\perp}  +\hat{H}^\perp+ \hat{H}_{\textrm{harm}}\right],\nonumber 
\end{align}

\noindent where $\hat{a}^{\dagger}_{i,\mathbf{r}^\perp}$ and $\hat{a}_{i,\mathbf{r}^\perp}$ are the bosonic creation and annihilation operators on lattice site $i$ and transverse position $\mathbf{r}^\perp$, $\hat{n}_{i,\mathbf{r}^\perp}$ is the corresponding number operator, $J/h\!=\!108(7)\,$Hz is the tunnel coupling, $U$ is the on-site interaction, $K\!=\!F_0 d$, $\hat{H}^\perp$ denotes the kinetic energy along the transverse direction and $\hat{H}_{\textrm{harm}}$ the 3D harmonic confinement. The Feshbach resonance of $^{39}$K at $403.4(7)\,$G enables us to work in the weakly-interacting regime at the scattering length $a_s\!=\!20 a_0$ (with $a_0$ the Bohr radius), where a mean-field approach is expected to be valid. The Feshbach resonance is crucial in order to find experimental parameter regimes, where parametric instabilities could be clearly identified in momentum-space images. Additional nonlinear effects rapidly counteract the exponential growth of the instabilities (\cite{Lellouch:2017it} and App.~\ref{sec:SI_inst_rates}), leaving only a small window of suitable parameters to observe it. 

In the non-interacting limit, Floquet theory predicts that the dynamics is well described by a time-independent Hamiltonian, with renormalized tunnel coupling $J_{\text{eff}}\!=\! J \mathcal{J}_0(\alpha)$~\cite{Lignier:2007du,Goldman:2014bva,Bukov:2015gu,Eckardt:2017hca}.
Here $\mathcal{J}_{\nu}$ is the $\nu$th-order Bessel function of the first kind and $\alpha\!=\!K/(\hbar\omega)$. Note, that $\alpha$ is independent of the modulation frequency $\omega$. The measurements are performed in a strong-driving regime, $1\!<\!\alpha\!<\!2$, where the effects of the drive are non-perturbative, while the minimum of the effective dispersion remains at $q^x\!=\!0$~\cite{Lignier:2007du,Michon:2017tw}. To avoid single-particle inter-band resonances~\cite{Weinberg:2015gc}, the modulation frequency is chosen well below the first single-particle bandgap of the static lattice $\Delta_{21}/h=41.6(5) \text{kHz}$. Indeed, we do not observe excitations to higher bands during the measurements, which would be visible in the TOF images (Fig.~\ref{Fig_2}a). 

In order to identify instabilities in the system, we monitor the appearance of collective excitations by measuring the momentum distribution of the atoms as a function of modulation time for various modulation frequencies and amplitudes. We find that after few modulation periods a small fraction of atoms is excited into additional momentum components that are distinct from the initial condensate at $q^x=0$ (Fig.~\ref{Fig_2}a). The amplitude of these modes grows as a function of the modulation time and excitation peaks eventually start to broaden after long modulation times (see $t=17\,T$ in Fig.~\ref{Fig_2}a) due to saturation and nonlinear effects (App.~\ref{sec:SI_inst_rates}).

We attribute these excitations to parametric instabilities that occur whenever a resonance exists between the energy associated with one drive quantum $\hbar \omega$ and the Bogoliubov spectrum $E_{\text{eff}}^B(\textbf{q})$ [see Eq.~(\ref{eq:Eav}) in App.~\ref{subsec:AnaRes}] of collective excitations~\cite{Kramer:2005dk,Creffield:2009cp,Lellouch:2017it}. In the presence of transverse modes there is no stable parameter regime, since the transverse kinetic energy has no upper bound. There is always a set of resonant excitations determined by the resonance condition 
\begin{equation}
\hbar \omega \!=\! E_{\text{eff}}^B(\textbf{q}_{\text{res}}), \quad \textbf{q}_{\text{res}}\!=\!(q^x_{\text{res}},\textbf{q}^{\perp}_{\text{res}}),
\end{equation}

\noindent which causes the system to be necessarily unstable. Each collective mode grows at a different rate $\Gamma_{\textbf{q}}$ and we denote the one associated with the dominating growth rate $\Gamma\!=\!\underset{{\textbf{q}}}{\text{max}}\ \Gamma_{\textbf{q}}$ the most unstable mode (``mum") with the corresponding momentum $\textbf{q}_{\text{mum}}$. 

For simplicity, we henceforth neglect the 3D harmonic confinement $\hat{H}_{\text{harm}}$ in our theoretical analysis and set the transverse kinetic term equal to a free-particle kinetic energy $\hat{H}^{\perp}=\sum_{j,\mathbf{q}^{\perp}} \frac{(\hbar \mathbf{q}^{\perp})^2}{2m} \hat{n}_{j,\mathbf{q}^{\perp}}$. For weak harmonic confinement, where the transverse modes are closely spaced, we expect this to be a good approximation. We note, however, that the harmonic confinement along the lattice axis largely modifies the stability of a strictly 1D system, where it prevents the existence of a true stable parameter regime, as the energy spectrum becomes unbounded (App.~\ref{sec:SI_trap}).

Following the approach developed in \cite{Creffield:2009cp,Lellouch:2017it} based on the time-dependent Bogoliubov-de-Gennes (BdG) equations of motion~\cite{supplements}, we find approximate analytic solutions for the momentum $\textbf{q}_{\text{mum}}$ and the growth rate $\Gamma$ of the most unstable mode (Ref.~\cite{Lellouch:2017it} and App.~\ref{subsec:AnaRes}). To lowest order in this perturbative treatment, there is a clear separation between the lattice and transverse degrees of freedom, resulting in two distinct regimes (Fig.~\ref{Fig_1}c):
\begin{enumerate}[(I)]
\item $\ \omega < \omega_{\text{sat}}$: $\quad q^x_{\text{mum}}<\pi/d \ $ and $\ |\mathbf{q}^{\perp}_{\text{mum}}|=0$,
\item $\ \omega > \omega_{\text{sat}}$: $\quad q^x_{\text{mum}}=\pi/d \ $ and $\ |\mathbf{q}^{\perp}_{\text{mum}}|>0$.
\end{enumerate}
In the first regime~(I), the modulation mainly couples to excitations along the lattice direction
\begin{equation}
q^x_{\text{mum}}\!=\!2\ \text{arcsin}\sqrt{\left(\sqrt{g^2+(\hbar\omega)^2}-g \right)/(4J_{\text{eff}})}/d \label{eq_qxmum},
\end{equation}
and the instability is not affected by the transverse modes ($|\mathbf{q}^{\perp}_{\text{mum}}|=0$); here $g=nU$ denotes the interaction energy and $n$ is the mean density (see App.~\ref{sec:SI_trap} for a discussion of the inhomogeneous density profile of a harmonically-trapped gas). When reaching the second regime~(II), in contrast, the transverse degrees of freedom dominate; $\mathbf{q}^{\perp}_{\text{mum}}$ becomes finite and grows according to 
\begin{equation}
\hbar^2 (\mathbf{q}^{\perp}_{\text{mum}})^2/(2m)\!=\!\sqrt{g^2+(\hbar\omega)^2}-g-4J_{\text{eff}}. \label{eq_perp}
\end{equation}


To quantify the position of the most unstable mode experimentally from the TOF images (Fig.~\ref{Fig_2}a), we subtract the mean initial condensate profile at $t=0$ from each individual TOF image, measured at $t>0$. A typical result is shown in Fig.~\ref{Fig_2}b (a more detailed description of the data analysis can be found in the Supplementary Material~\cite{supplements}). To characterize the excitations along the lattice direction $q^x_{\text{mum}}$, we integrate the 2D difference profile perpendicular to the lattice axis (lower panels in figure~\ref{Fig_2}b). The resulting 1D profile is divided into two parts excluding the negative part at small $|q^x|$ (purple), which arises due to the small depletion of the condensate. We determine the position $q^x_{\text{max}}$ and amplitude $A^x$ of the excitation peaks by fitting asymmetric Lorentzian functions to each peak [Eq.~(\ref{eq:asymLorentz}) in the Supplementary Material].  

The transverse momentum component $\mathbf{q}^{\perp}_{\text{mum}}$ of the most unstable mode is masked by the initial momentum spread of the condensate and the width of the parametric resonance. Nevertheless, it manifests itself in a broadening, which we monitor by extracting the transverse width $\Delta q^y$ of the excitation peaks: We integrate the 2D difference profile along the lattice direction using a region of interest that contains only the left or the right peak. We fit the resulting 1D profiles (left and right panel in figure~\ref{Fig_2}b) with a symmetric Lorentzian and extract the full width at half maximum.
The fit parameters obtained for the left and right peak are then averaged over all images for each modulation time $t>0$.

\begin{figure}
\centering
\includegraphics{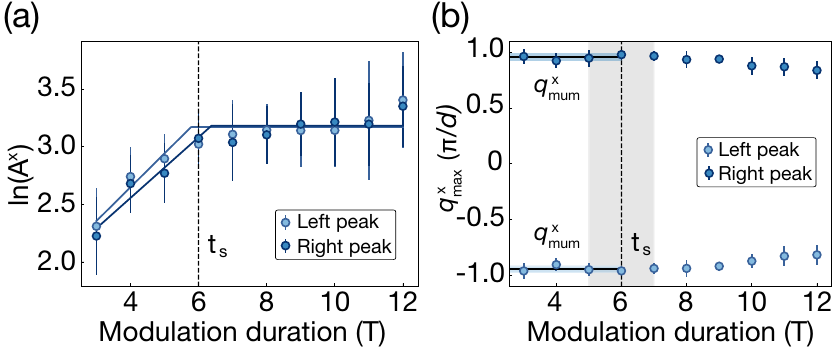}
\vspace{-0.cm} \caption{Time-resolved growth of the most unstable mode for $\alpha=1.78$ and $\omega/(2\pi) = 720 \text{Hz}$. (a) Logarithmic plot of the peak amplitude $A^x$ as a function of modulation time. The solid lines are fits to the data to extract the end of the short-time regime $t_s$ (black dashed line). (b) Position of the peak maximum $q^x_{\text{max}}$ along $q^x$ as a function of modulation time. The most unstable mode $q^x_{\text{mum}}$ is defined as the weighted mean of all $q^x_{\text{max}}$ for $t \leq t_s$. Its value is shown as the solid black line; the shaded blue bars denote its standard error of the weighted mean. The shaded gray area defines the range of hold times over which the transverse width $\Delta q^y$ is averaged. Each data point is an average over $\sim 10$ individual experimental realizations, error bars indicate the standard deviation. To minimize systematic deviations in the bandmapped images, we average $\lessim\,5$ realizations for a modulation phase $\varphi=0$ and $\lessim\,5$ with $\varphi=\pi$ (Fig.~\ref{Fig_asymmetry}).}
\label{Fig_3}
\end{figure}

In Fig.~\ref{Fig_3} we show a typical time series for the position $q^x_{\text{max}}$ and amplitude $A^x$ of the dominant mode (full data set is shown in the Supplementary Material~\cite{supplements}). While there are remnants of a moderate exponential growth at short times $t<t_s$ we predominantly find an explicitly time-dependent growth rate. This is due to incoherent processes caused by interactions between collective excitations and between the excited modes and the condensate~\cite{Lellouch:2017it}. This behavior is confirmed by our numerical simulations beyond BdG theory discussed in App.~\ref{sec:SI_inst_rates}. In order to quantitatively compare our experimental data with analytical expressions obtained using Bogoliubov theory we restrict the analysis to a small time window $t<t_s$ before saturation and nonlinear effects start to dominate. This is motivated by the good agreement between numerical simulations and analytical formulas for short times (Fig.~\ref{Fig_9} in App.~\ref{sec:SI_inst_rates}). The parameter $t_s$ is evaluated by fitting a piecewise function consisting of a linear and a constant part to $\text{ln}(A^x)$ individually for the left and right excitation peak (Fig.~\ref{Fig_3}a) and averaging the two results~\cite{supplements}. The corresponding growth rates (for $t<t_s$) are in good agreement with Bogoliubov theory (Fig.~\ref{Fig_10} in App.~\ref{sec:SI_inst_rates}), validating this approach. The position of the most unstable mode $q^x_{\text{mum}}$ is then defined as the average over the peak positions $q^x_{\text{max}}$ for all modulation times $t<t_s$. We observe a decrease of $q^x_{\text{max}}$ at modulation times $t>t_s$ indicating the onset of additional scattering events not captured by Bogoliubov theory. In order to study the behavior along the transverse direction, we analyze the peak width $\Delta q^y$ at the end of the short time regime, where the amount of transverse excitations is expected to be maximal (full data set is shown in the Supplementary Material~\cite{supplements}). In order to reduce statistical errors, we average $\Delta q^y$ over modulation times between $t_s -1$ and $t_s +1$ in units of the modulation period (Fig.~\ref{Fig_3}b) to obtain the experimental value for the width $\Delta q^y_{\text{lin}}$~\cite{supplements}.

\begin{figure}[t!]
\includegraphics{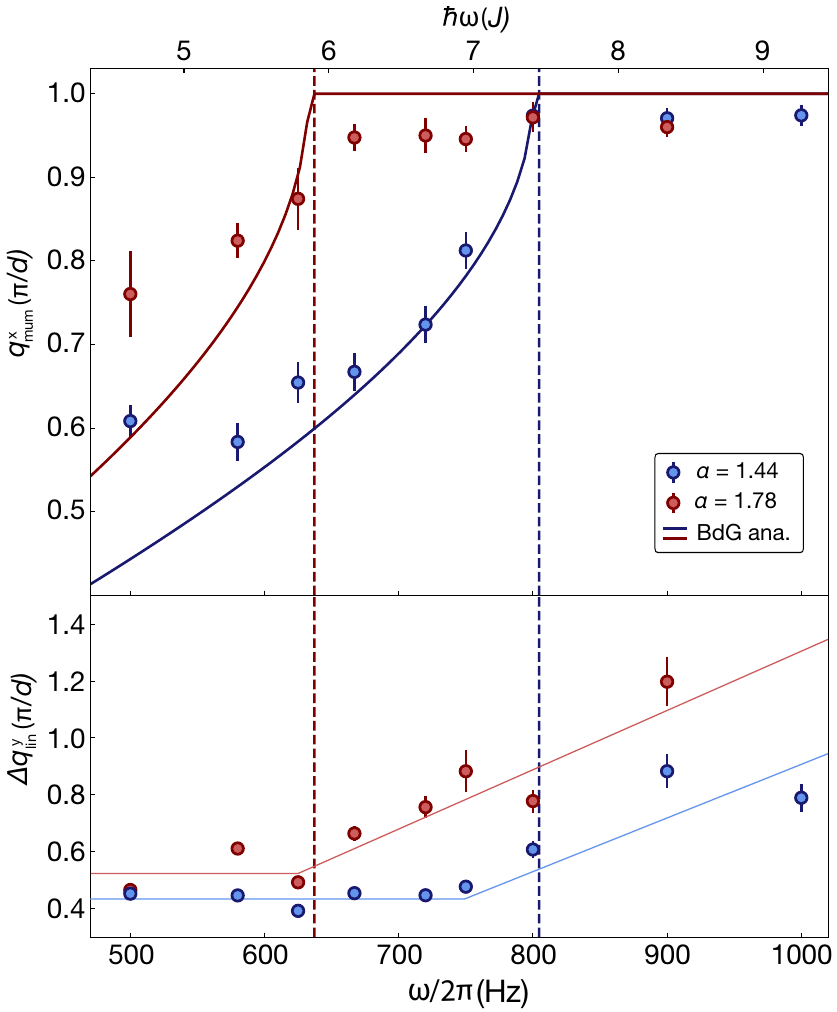}
\vspace{-0.cm} \caption{Position of the most unstable mode and influence of transverse modes. 
Upper panel: Position of the most unstable mode $q^x_{\text{mum}}$ as a function of modulation frequency for $\alpha=1.44$ and $\alpha=1.78$. The error bars denote the standard error of the mean. The solid lines show Eq.~(\ref{eq_qxmum}) for $g=11.5\,J$. The vertical dashed lines mark the corresponding saturation frequency. Lower panel: Transverse width $\Delta q^y_{\text{lin}}(\omega)$. Error bars denote the standard error of the weighted mean. The thin lines are guides to the eye.}
\label{Fig_4}
\end{figure}

The position of the most unstable mode along the lattice axis $q^x_{\text{mum}}$ and the transverse width $\Delta q^y_{\text{lin}}$ are measured for various modulation parameters around the saturation frequency, where we expect a crossover between parametric instabilities dominated by excitations along the lattice axis [regime (I)] and those that are facilitated by the presence of transverse modes [regime (II)]. The results are shown in Fig.~\ref{Fig_4}.

We find that $q^x_{\text{mum}}$ indeed increases with $\omega$ until it saturates at $q^x_{\text{mum}}\!\approx\!\pi/d$, the edge of the Brillouin zone, at a frequency $\omega_{\text{sat}}$ that depends on the driving amplitude $\alpha$ (Fig.~\ref{Fig_4} upper panel). The small deviation of the measured positions from $\pi/d$ is mainly due to the short TOF used in the experiment~\cite{supplements}. The saturation frequency $\omega_{\text{sat}}$ calculated from Eq.~(\ref{eq_qxmum}) matches the experimental data well for an interaction parameter $g=11.5\,J$. From our measured in-situ density profiles we obtain an interaction parameter $g_{\text{max}}\approx8\,J$ in the center of the trap. 
The deviation is most likely due to a systematic uncertainty in the atom number calibration~\cite{supplements}. At the same time, we observe that the width $\Delta q^y_{\text{lin}}(\omega)$ transverse to the lattice axis starts to increase with frequency for $\omega>\omega_{\text{sat}}(\alpha)$, simultaneously with the saturation of $q^x_{\text{mum}}$ at the BZ edge (Fig.~\ref{Fig_4} lower panel), as expected from lowest-order perturbation theory Eqs.~(\ref{eq_qxmum})-(\ref{eq_perp}). Moreover, the transition point $\omega_{\text{sat}}(\alpha)$ decreases for larger driving parameters $\alpha$, in line with the reduction of $J_\text{eff}$.

We further observe that the shape of the $q^x_{\text{mum}}$-curve differs from the theoretical prediction for low modulation frequencies in particular for the large modulation amplitude $\alpha=1.78$. We attribute this deviation mainly to the following effects: First, the short-time window, during which we can clearly identify the most unstable mode strongly depends on the modulation parameters. The time $t_s$, which marks the end of the short-time window, decreases with decreasing modulation frequency and increasing modulation amplitude (Fig.~\ref{Fig_S1}). This complicates the identification of the peak maxima. If $t_s$ becomes too small, there is not enough time for the most unstable mode to dominate over the other excited modes. 
Second, the initial width of the momentum distribution of the condensate at $t=0$ poses a fundamental limitation that prevents us from measuring positions smaller than $\approx 0.4 \pi/d$. Moreover, for large modulation amplitudes higher-order corrections to the analytical formulas Eqs~(\ref{eq_qxmum})-(\ref{eq_perp}) become relevant (App.~\ref{subsec:AnaRes}).

In summary, we have demonstrated the first direct evidence for parametric instabilities in shaken optical lattices via momentum-resolved measurements. Our experiments were performed with a large 3D system of weakly-interacting bosons, where exact numerical calculations including the harmonic trap are not feasible. Tuning the scattering length with a Feshbach resonance we were able to identify and address experimental parameter regimes, where the short-time dynamics is well-described by BdG theory. While the instability rates are time-dependent due to competing processes that dominate on different timescales, the momentum of the most unstable mode turned out to be a reliable observable. The obtained results are in agreement with the analytic approach derived in Ref.~\cite{Lellouch:2017it}, which enables the development of an intuitive understanding and allows us to identify stable parameter regimes in driven lattice models from simple energetic arguments. We were able to verify the existence of key bottlenecks in current experimental settings with weak transverse confinement~\cite{Struck:2011is,Aidelsburger:2011hl,Atala:2014uo,Ha:2015vu,Aidelsburger:2015hm,Kennedy:2015dw,Li:2017ha} that need to be overcome by freezing the transverse degrees of freedom and generating a box-type longitudinal confinement~\cite{Corman:2014cm,Navon:2015jd}. Parametric instabilities can indeed lead to a depletion of  the condensate, where subsequent scattering events result in large heating rates. Our results are of strong interest for future experiments based on Floquet engineering \cite{Eckardt:2017hca}, as they indicate the necessity to engineer full 3D lattice systems, where stable regimes can be found \cite{Creffield:2009cp,Lellouch:2017it}. Parametric resonances are expected to be present whenever the BdG equations of motion include time-periodic features, and hence may turn out to play an important role in a wide family of Floquet-engineered systems, such as periodically-driven superfluids~\cite{Eckardt:2017hca} and superconductors~\cite{Babadi:2017dc}, photonic devices~\cite{Peano:2016ds,GonzalezTudela:2015gd}, but also in the context of cosmology~\cite{Berges:2014bl}.

We note that during the completion of this work, short-time heating rates, which are expected to be dominated by parametric instabilities, have been investigated in modulated 2D lattices~\cite{refPorto}.


We thank T. Boulier, C. Braun, M. Cheneau, C. Chin, T. Esslinger, L. Fallani, D. Gu\'ery-Odelin, A.~Polkovnikov, T. Porto, M.~Reitter, L. Tarruell, and D.~Sels for insightful discussions. The work in Munich was supported by the Deutsche Forschungsgemeinschaft
(FOR2414 Grant No. BL 574/17-1), the European Commission (UQUAM Grant No. 5319278, AQuS), the Nanosystems Initiative Munich (NIM) Grant No. EXC4 and by the DeutscheForschungsgemeinschaft (DFG, German Research Foundation) under Germany's Excellence Strategy -- EXC-2111 -- 39081486. The work in Brussels was financed by the FRS-FNRS (Belgium) and the TopoCold ERC Starting Grant. U.~S. acknowledges support from the EPSRC Programme Grant DesOEQ (EP/P009565/1). M.~B. acknowledges support from the Emergent Phenomena in Quantum Systems initiative of the Gordon and Betty Moore Foundation, and the U.S. Department of Energy, Office of Science, Office of Advanced Scientific Computing Research, Quantum Algorithm Teams Program. E.~D. was supported by Harvard-MIT CUA, NSF Grant No. DMR-1308435, AFOSR Quantum Simulation MURI, AFOSR-MURI Photonic Quantum Matter (award FA95501610323). We used QuSpin~\cite{Weinberg:2017ch,Weinberg:2018vq} to perform the numerical simulations. The authors are pleased to acknowledge that the computational work reported on in this paper was performed on the Shared Computing Cluster which is administered by Boston University's Research Computing Services.


%

\cleardoublepage

\begin{appendix}

\section{\label{subsec:AnaRes} Analytical treatment of parametric instabilities within the Bogoliubov approximation}

Here, we would like to recall the analytical method developed in Ref.~\cite{Creffield:2009cp} to extract the instability properties of the system within the Bogoliubov approximation. As mentioned in the main text, for simplicity, we neglect the harmonic confinement, so that the transverse kinetic energy is determined by the free-particle dispersion relation. It has been shown that the Bogoliubov equations of motion [Eq.~(\ref{eq:BdGE_EOM}) in Sect.~\ref{subsec:Bog_lin} of the Supplementary Material~\cite{supplements}], can be mapped to a parametric oscillator model~\cite{Landau:1969,Bukov:2015dv}, a seminal model of periodically-driven harmonic oscillator known to display parametric instabilities as soon as the drive frequency approaches twice the natural frequency.
To see that, one should perform a series of suitable changes of basis and reference frames~\cite{Lellouch:2017it}. Applying the Rotating Wave Approximation (RWA), we keep the leading-order harmonic and recast the Bogoliubov equations into the form:
\begin{widetext}
	\begin{align}
		&i \hbar \partial_t \! \left( \begin{matrix} \tilde{u}'_\qbf  \\ \tilde{v}'_\qbf \end{matrix} \right)=\biggl[ \Eav(\qbf)\hat{\mathbf{1}}
		+\!\dfrac{A_\qbf\Eav(\qbf)}{2}\left( \begin{matrix} 0 & \cos(2\omega t)\mathrm e^{-2i\Eav(\qbf)t/\hbar}        \\ -\cos(2\omega t)\mathrm e^{2i\Eav(\qbf)t/\hbar} & 0 \end{matrix} \right)\biggr]\! \left( \begin{matrix} \tilde{u}'_\qbf  \\ \tilde{v}'_\qbf \end{matrix} \right),
		\label{eq:BdGEPO}
	\end{align}
	where $\hat{\mathbf{1}}$ is the identity matrix,
	\begin{align}
		\Eav(\mathbf{q})=&\sqrt{\left(4|\Jeff|\sin^2 (q^x d/2) + \left(\hbar \qp\right)^2/2m \right) \left(4|\Jeff|\sin^2 (q^xd/2)+\left(\hbar \qp\right)^2/2m +2g\right)},\label{eq:Eav}		
	\end{align}
\end{widetext}
denotes the effective (time-averaged) Bogoliubov dispersion, and we introduced the amplitude
\begin{equation}
A_\qbf=16J \mathcal{J}_2(\alpha)\sin^2(q^x d/2) \frac{g}{[\Eav(\qbf)]^2}.
\label{eq:amp}
\end{equation}
For the sake of simplicity, we have here restricted ourselves to the dominant harmonic of the drive, and dropped all terms in Eq.~(\ref{eq:BdGEPO}) that are irrelevant regarding the occurrence of instabilities -- a simplification that was rigorously established in Ref.~\cite{Lellouch:2017it}. 

\begin{figure}[t!]
	\includegraphics{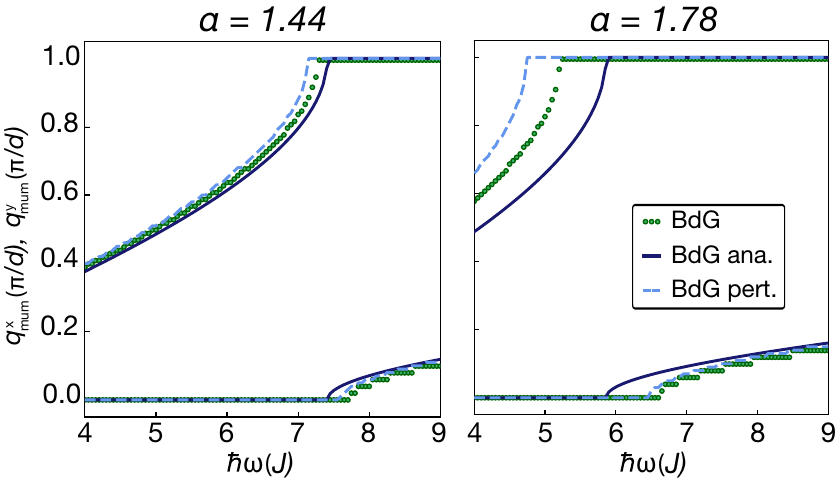}
	\vspace{-0.cm} \caption{Comparison of the perturbative analytic treatment calculated to zeroth [solid blue line, Eq.~(\ref{eq_qxmum}) and (\ref{eq_perp})] and second order (dashed blue line) with numerical simulations of the exact BdG equations (green dots) for $g/J=11.5$ and two modulation amplitudes: $\alpha=1.44$ (left) and $\alpha=1.78$ (right). The steps in the numerical results are due to the discretization in momentum space.
		\label{Fig_5}
	}
\end{figure}

\paragraph*{Perturbation theory---}As can be seen from a RWA treatment of Eq.~(\ref{eq:BdGEPO}), each momentum mode $\qbf$ will display a dynamical instability (characterized by an exponential growth of its population), whenever the drive frequency $\omega$ approximately matches its effective Bogoliubov energy $\Eav(\mathbf{q})$, i.e., $\hbar\omega\!\approx\!\Eav(\qbf)$. The analytical method to extract the associated instability rate is detailed in Refs.~\cite{Landau:1969,Lellouch:2017it} and relies on a perturbation theory formulated in both $A_\qbf$ (the small parameter of the expansion, which is assumed to be smaller than $1$ for the expansion to converge), and the detuning from the resonance $\delta_{\textbf{q}}\equiv \hbar\omega-\Eav(\qbf)$. In brief, the findings are the following:\\

\begin{enumerate}
\item Zeroth order: the instability occurs only on resonance, i.e., $\hbar \omega=\Eav(\qbf)$ and the instability rate is given by
	\begin{align}
	\Gamma^*_{\textbf{q}}&=\frac{A_{\textbf{q}}\Eav(\textbf{q})}{4h} \nonumber\\
	&=	\frac{4J}{h} \mathcal{J}_2(\alpha)\sin^2(q^xd/2)\dfrac{g}{\hbar\omega},
\label{eq:sres}	
	\end{align}
for all $\qbf$ fulfilling the resonance condition $\hbar \omega=\Eav(\qbf)$ and is zero for all other modes. Interestingly, not all resonant modes necessarily have the same instability rate. We focus on the most unstable mode, which has the largest rate $\Gamma = \underset{{\textbf{q}}}{\text{max}} \ \Gamma^*_{\textbf{q}}$, which results in the following rates for the two regimes introduced in the main text:
\begin{align}
\qquad \textrm{(I)} \ &\omega < \omega_{\text{sat}}: \nonumber\\
&\Gamma = \frac{1}{h}\left(\sqrt{g^2+(\hbar\omega)^2}-g\right) \left|\dfrac{\mathcal{J}_2(\alpha)}{\mathcal{J}_0(\alpha)}\right|\dfrac{g}{\hbar\omega},\label{eq:gamma_BdG1} \\
\qquad \textrm{(II)} \ &\omega > \omega_{\text{sat}}: \nonumber \\
& \Gamma = \frac{4J}{h}\left|\mathcal{J}_2(\alpha)\right|\dfrac{g}{\hbar\omega}. \label{eq:gamma_BdG2}
\end{align}
The corresponding momentum $\qbf_{\text{mum}}$ of the most unstable mode to lowest order in this pertubative treatment is defined in Eq.~(\ref{eq_qxmum}) and (\ref{eq_perp}) in the main text.

\item First order: we find that instability does not only occur on resonance but still arises in a finite window around the resonance point. The associated instability rate is given by 
	\begin{equation}
	\Gamma_{\textbf{q}}=\dfrac{A_{\textbf{q}}\Eav(\textbf{q})}{4h}\sqrt{1-\left(\frac{\delta_{\textbf{q}}}{A_\qbf\Eav(\qbf)}\right)^2},
	\label{eq:s1ord}	
	\end{equation}
if the argument of the square root is positive, and $\Gamma_{\textbf{q}}\equiv 0$ otherwise. It is maximal at resonance and decreases  with the distance from resonance, until it vanishes at the edges of the instability domain~\cite{Landau:1969,Lellouch:2017it}.\\
\item Second order: We no longer have explicit analytical expressions, but we find that $\Gamma_{\textbf{q}}$ is the solution of an implicit equation which can be numerically solved~\cite{Landau:1969,Lellouch:2017it}. While the width of the resonance is unaffected, we obtain that the instability rate is no longer maximal on resonance, but that the maximum instability point is slightly shifted from the resonance.
\end{enumerate}

In Fig.~\ref{Fig_5} we show a comparison between the analytic formulas Eq.~(\ref{eq_qxmum}) and (\ref{eq_perp}), the perturbation theory up to second order and a numerical simulation of the exact time-dependent BdG equations Eq.~(\ref{eq:BdGE_EOM}). We find that, for large modulation amplitudes, higher-order corrections become important because $A_\qbf$ [Eq.~(\ref{eq:amp})] is not a small parameter any more; more specifically $\mathcal{J}_2(\alpha)$ is no longer small compared to the effective Bogoliubov dispersion $\Eav(\textbf{q})$. We believe that this explains why the measured $q_{\text{mum}}^x$-curve in Fig.~\ref{Fig_4} deviates more from the zero-order BdG theory for large modulations amplitudes, i.e. $\alpha=1.78$, as compared to the rather good agreement observed for $\alpha=1.44$. We stress that the BdG analysis we present here does not rely on the inverse-frequency expansion, which is routinely used to determine the effective Hamiltonian in the context of a time-periodic Schr\"odinger equation. 

\section{\label{sec:SI_trap}Harmonic confinement along the lattice direction}

In this section, we analyze modifications to the translationally-invariant Bogoliubov-de-Gennes (BdG) theory [Sect.~\ref{subsec:Bog_lin} in the Supplementary Material~\cite{supplements}] due to the harmonic trap along the lattice axis. We constrain the discussion to 1D systems for simplicity.

\subsection{\label{sec:SI_trap_energy}Unbounded energy spectrum}

\begin{figure}[t!]
	\includegraphics{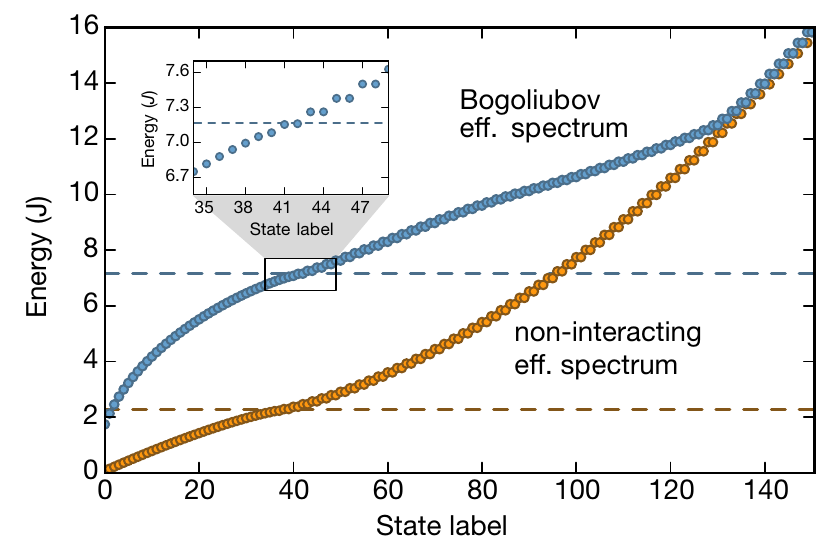}
	\vspace{-0.cm} \caption{Energy spectrum of the non-interacting time-averaged Hamiltonian (orange dots) and corresponding effective Bogoliubov dispersion (blue dots) for $\alpha=1.4$ and $g_\mathrm{max}=10.2J$ in a harmonic trap with $\omega_x = 0.26~J/\hbar$. The dashed lines indicate the respective bandwidths in the homogenous case, i.e. $\hbar\omega_\mathrm{sat}=2.2\,J$ (orange) and $\hbar\omega_\mathrm{sat}=7.17\,J$ (blue) according to Eq.~(\ref{eq:omega_sat_trap}). Inset: As the energy reaches the homogenous bandwidth, the spectrum becomes quasidegenerate.
		\label{Fig_6}}
\end{figure}

In most optical lattice experiments there is a harmonic confinement along the lattice axis, and therefore the energy spectrum of the system is no longer bounded from above. Figure~\ref{Fig_6} (orange dots) shows the energy spectrum of the effective time-averaged non-interacting Hamiltonian in the presence of the harmonic trap, and the corresponding drive-renormalized Bogoliubov dispersion (blue dots). Details about the numerical simulations of the inhomogeneous system can be found in Sec.~\ref{sec:numericsImhom} of the Supplementary Material~\cite{supplements}. Notice the occurrence of states above the Bogoliubov bandwidth (blue dashed line), which can be excited resonantly by the drive. In momentum space, these states occupy modes in the vicinity of $q^x=\pi/d$, and typically have weight over a finite range of momenta. Therefore, the presence of a weak harmonic confinement allows the system to absorb energy even for $\omega > \omega_\mathrm{sat}$, where unconfined systems are shown to be stable~\cite{Lellouch:2017it}. As a result,  we do not expect a truly stable parameter regime to exist in harmonically confined systems, even in the absence of transverse modes. 

\subsection{\label{sec:SI_trap_omegasat}Modified saturation frequency $\omega_\mathrm{sat}$}

The expressions for the saturation frequency $\omega_\mathrm{sat}$ and the effective bandwidth $W_\mathrm{eff}$, derived using translationally-invariant Bogoliubov theory, depend on the density-renormalized interaction parameter $g$. To zeroth-order (App.~\ref{subsec:AnaRes}) the effective bandwidth for the 1D lattice is obtained from Eq.~(\ref{eq:Eav}) by setting $\qbf^\perp=0$. In the presence of a harmonic confinement, the condensate profile obeys Thomas-Fermi theory and is no longer uniform, which induces a position dependence in $g\to g(x)$. Since the energy spectrum is unbounded, there is no natural energy bandwidth to use. Hence, we need to determine the appropriate value for $g$ used to compute the saturation frequency $\omega_\mathrm{sat}$.

We observe a correlation between the energy at which the states in the Bogoliubov spectrum rapidly become quasidegenerate [cf.~Fig.~\ref{Fig_6} inset], and the energy scale $\omega_\mathrm{sat}(g_\mathrm{max})$ with $g_\mathrm{max}=\underset{{x}}{\text{max}}\ g(x)$, where

\begin{equation}
\hbar \omega_\mathrm{sat}(g,\alpha)=\sqrt{4J_\mathrm{eff}(\alpha)(4J_\mathrm{eff}(\alpha) + 2 g ) }.
\label{eq:omega_sat_trap}
\end{equation}
The same scale also coincides with the energy, starting from which the Bogoliubov states attain a significant occupation of the $q^x=\pi/d$ mode. This suggests that, in the presence of a harmonic trap, the maximum value $g_\mathrm{max}$ can be used to estimate the saturation frequency $\omega_\mathrm{sat}$.

\begin{figure}[t!]
	\includegraphics{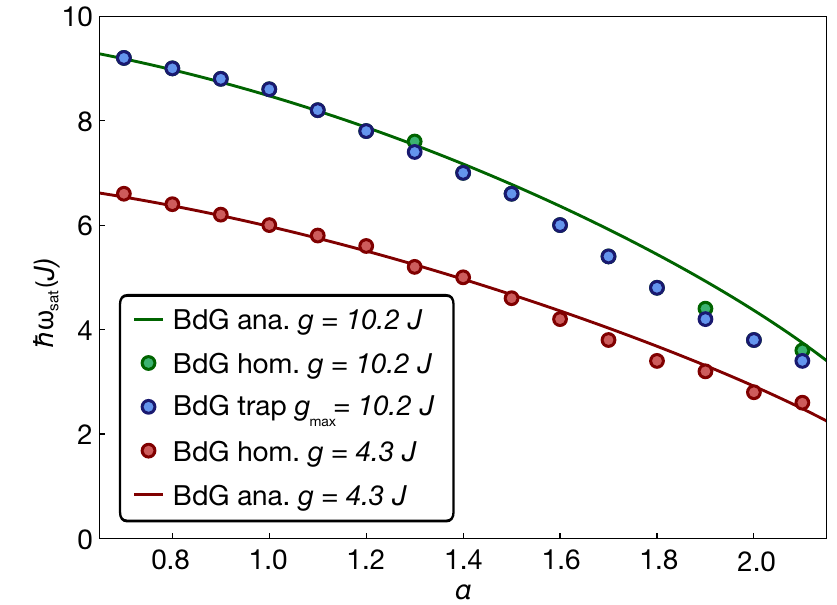}
	\vspace{-0.cm} 
	\caption{Numerical simulation of the saturation frequency $\omega_{\text{sat}}$ vs.~$\alpha$ [dots] compared to the BdG prediction [solid line] from Eq.~(\ref{eq:omega_sat_trap}).
		The presence of a harmonic trap does not affect the saturation frequency significantly. 
		A slight deviation from the BdG prediction is observed compared to the numerical simulation for large values of $\alpha\sim 2$ and sufficiently strong effective interaction $g/J$.
		The trapping frequency is $\omega_x=0.26\,J/\hbar$, the system size is $L_x=201\,d$ and the atom number is $N_0=1000$.
		\label{Fig_7}}
\end{figure}

To test this conjecture, we performed numerical BdG simulations of 1D lattices comparing the position of the most unstable mode with and without harmonic confinement. Indeed, we find a qualitative behavior similar to the homogeneous system, i.e.,~the momentum $q^x_{\text{mum}}(\omega)$ increases with $\omega$ until it saturates at $q^x_{\text{mum}}\!\approx\!\pi/d$. 
We can also quantitatively compare the $\alpha$-dependence of the saturation frequency $\omega_\mathrm{sat}(\alpha)$ for the homogeneous and trapped systems as follows: (1) we fix a trap frequency and determine the value of $g_\mathrm{max}=U\underset{{x}}{\text{max}}\,n_\mathrm{TF}(x)$ from the Thomas-Fermi profile of the condensate wavefunction $n_\mathrm{TF}(x)$. (2) we simulate the dynamics of a homogeneous system using the same value of $g$. If Eq.~\eqref{eq:omega_sat_trap} for $g=g_\mathrm{max}$ provides a correct description, this procedure  will result in the same value of $\omega_\mathrm{sat}$ for the trapped and homogeneous simulations by construction. The values of $\omega_\mathrm{sat}(\alpha)$ are extracted numerically from the $q^x_\mathrm{mum}$ vs.~$\omega$ curves for every fixed value of $\alpha$, as follows:
(i) we compute numerically the momentum distribution profile of the Bogoliubov modes as a function of momentum $q^x$ and time $t$, evolved under the periodic drive. We do this for a grid of various $\alpha$ and $\omega$ points.
(ii) we evolve the system for $20$ driving cycles which is enough to single out the most unstable mode that grows exponentially according to the BdG equations~\cite{supplements}. 
(iii) we extract the fastest growing mode $q^x_\mathrm{mum}$ from the latest time slice for every point on the $(\alpha,\omega)$ grid. 
(iv) for every fixed $\alpha$, we determine the saturation frequency, by finding the frequency for which $q^x_\mathrm{mum}$ reaches $\pi/d$ for the first time upon increasing $\omega$.  

Figure~\ref{Fig_7} shows that the behavior of the saturation frequency $\omega_{\text{sat}}(\alpha)$ agrees with analytic predictions of Eq.~\eqref{eq:omega_sat_trap} for $g=g_\mathrm{max}$. We observe that the agreement gradually becomes limited in the regime of large $\alpha$ where the effective kinetic energy of the system is parametrically reduced for strong interactions $g$. This regime of large $\alpha$ and large $g/J$ is precisely where higher-order corrections to Eq.~\ref{eq:amp} become pronounced (see the discussion in App.~\ref{subsec:AnaRes}). This explains the observed mismatch in the $g/J=10.2$ curves at large $\alpha$ in Fig.~\ref{Fig_7}.

\subsection{Modified BdG instability rates}

\begin{figure}[t!]
	\includegraphics{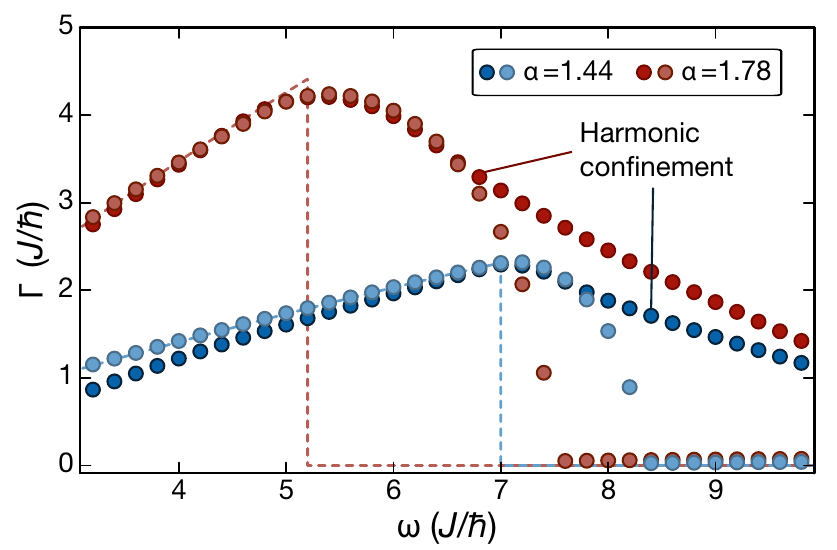}
	\vspace{-0.cm} \caption{Numerical simulations of the instability rate in homogeneous 1D lattices (light red and light blue, $g=10.2\,J$) in comparison with the rates in 1D lattices with harmonic confinement (dark red and dark blue, $\omega_x\!=\!0.26\,J/\hbar$ and $g_{\text{max}}=10.2\,J$) for two different driving parameters $\alpha=1.4$ (blue) and $\alpha=1.8$ (red). The system parameters are the same as for the simulations in Fig.~\ref{Fig_7}, i.e.,~$L_x\!=\!201\,d$ and $N_0\!=\!1000$. The instability rate was extracted from an exponential fit to the numerical data from the last five out of twenty-four driving cycles of evolution, to make sure the maximally unstable mode dominates. The dashed vertical lines show the BdG predictions for $\omega_\mathrm{sat}$ at $g=10.2\,J$.
		\label{Fig_8}
	}
\end{figure}

In Fig.~\ref{Fig_8} we show numerically evaluated instability rates of 1D lattice models using BdG simulations. We compare the drive frequency dependence for the inhomogeneous vs.~homogeneous lattices. 
Ideally, a homogeneous 1D system becomes stable for $\omega > \omega_{\text{sat}}$, since there are no states above the lattice bandwidth~\cite{WidthTest}.
The simulation reveals that the presence of the harmonic trap does not lead to any further shift in the position of $\omega_\text{sat}$, provided one compares a homogeneous system of interaction strength $g$ to a harmonically confined system with $g_\mathrm{max}=g$.
In the latter case, $\omega_{\text{sat}}$ is approximately given by the Bogoliubov bandwidth evaluated at $g_{\text{max}}$ [Fig.~\ref{Fig_7}]. 
For larger modulation frequencies $\omega$, the trapped system displays a distinctly different behavior compared to the homogeneous system. As already anticipated, due to the unbounded nature of the energy spectrum, there is no true stable (i.e.~$\Gamma\equiv0$) parameter regime and the confined system can always absorb energy via resonant processes. Nonetheless, we find a decrease of the instability rates with increasing drive frequency. 

Due to lack of translation invariance, it is not feasible to carry out an analytical calculation to determine the precise decay law $\Gamma(\omega)$ for $\omega>\omega_\mathrm{sat}$ in the presence of the harmonic trap. However, assuming the Local Density Approximation for a weak-enough harmonic confinement and keeping in mind the unbounded structure of the spectrum, we can apply Eqs~\eqref{eq:gamma_BdG1}-\eqref{eq:gamma_BdG2} in all spatial regions of approximately constant density, according to which $\Gamma\propto \omega^{-1}$ for $\omega>\omega_\mathrm{sat}$; while this does not predict the exact functional form of $\Gamma(\omega)$ observed in Fig.~\ref{Fig_8}, it presents an approximate argument for the observed rate decay. 

As a result, for trapped systems one can recover an approximately stable regime, where the instability rates are small compared to the duration of the experiment. An alternative way to mitigate the problem of non-vanishing rates at large drive frequencies in experiments could be provided by the use of uniform box traps~\cite{Corman:2014cm,Navon:2015jd}.

\section{\label{sec:SI_inst_rates}Instability rates}

Besides the position of the most unstable mode discussed in the main text, a characteristic property of parametric instabilities are the associated instability rates $\Gamma$, which after sufficiently long times are dominated by the growth of the occupation of the most unstable mode $n_{q_{\text{mum}}}$. Here, we present an experimental and numerical study of these rates.

\subsection{\label{sec:SI_inst_rates_num}Influence of mode coupling}

In order to provide more insight on the time-dependence of the instability rates found in the experiment (Fig.~\ref{Fig_3}a and Fig.~\ref{Fig_S1} in the Supplementary Material~\cite{supplements}), we performed numerical simulations (see~\cite{supplements} for details) on a homogeneous hybrid (translationally-invariant) 2D system~\cite{Note1}, composed of one lattice and one continuum direction, based on three different approximation methods~\cite{supplements}: (i) The linearized BdG equations, which capture the parametric instability at short times, i.e.,~before saturation effects, such as particle-number conservation, and non-linear effects associated with the Gross-Pitaevskii equation, become significant. (ii) The weak-coupling conserving approximation (WCCA)~\cite{Bukov:2015dv}, where particle-number conservation is restored, and which couples the condensate mode to the excitations to leading order in the interaction strength $U$. This method keeps track of the number of atoms scattered into finite-momentum modes, as well as the back action of the Bogoliubov quasiparticles onto the condensate. The WCCA, however, does not capture collisions between quasiparticles. Hence, it does not offer any insight on the thermalization dynamics at longer times, during which the system heats up steadily to an infinite-temperature state. The BdG and WCCA approaches have already been compared in Ref.~\cite{Lellouch:2017it} and their validity at short times is further confirmed in this work by comparing their predictions to a third approach: (iii) the Truncated Wigner approximation (TWA), which produces thermalizing dynamics and, even though quantum effects are only partially captured in this semi-classical approximation, it has recently been demonstrated that classical Floquet systems thermalize in a very similar way to quantum models~\cite{Howell:2018uy,Notarnicola:2018ft,Rajak:2018ta}.

\begin{figure}[t!]
\includegraphics{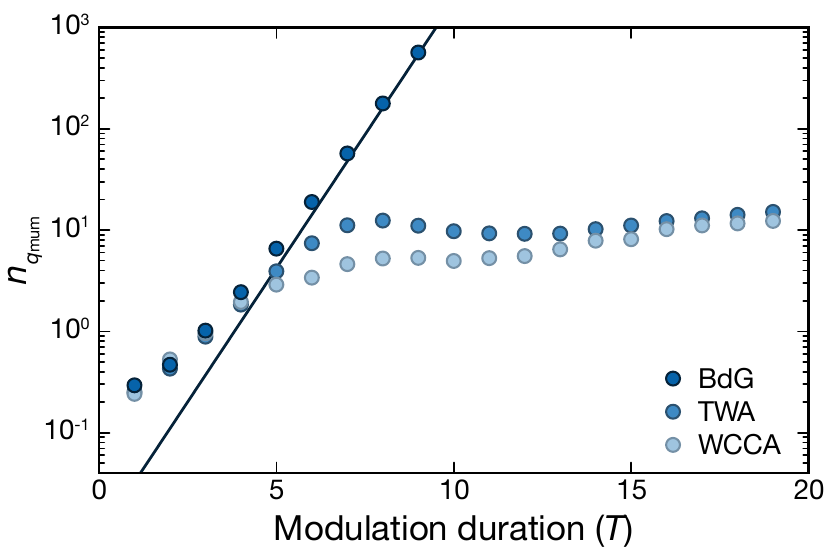}
\vspace{-0.cm} \caption{Numerical simulations of the occupation of the most unstable mode $n_{q_{\text{mum}}}$ for a homogeneous 2D system (1D lattice and one continuous direction) for $g=9.52\,J$, $\hbar\omega\!=\!9.25\,J$ and $\alpha=1.44$. The solid line displays the rate obtained from the analytic formulas in Eqs~\eqref{eq:gamma_BdG1}-\eqref{eq:gamma_BdG2}, which is in agreement with the BdG simulations (dark blue) for $t\gtrsim 5T$. The TWA (blue) and WCCA (light blue) partially capture additional non-linear effects, which result in a time-dependent instability rate. 
\label{Fig_9}
}
\end{figure}

Figure~\ref{Fig_9} shows the time evolution of the occupation of the most unstable mode. In both, the WCCA and TWA simulations, the obtained curves directly reflect the condensate depletion dynamics. At short times, we find an agreement between the three theoretical approaches, which further improves as one decreases the on-site interaction strength $U$, while keeping $g=nU$ fixed. At later times, the three approximations exhibit different behaviors: due to the lack of particle conservation, the BdG curves grow exponentially in an unphysical and indefinite manner. In contrast, the WCCA and TWA curves show clear manifestations of saturation effects, indicating that the instability rate is a truly time-dependent physical quantity. After an intermediate transient, all momentum modes are equally populated, which is expected for an infinite-temperature state that is reached at long times. 
These results offer a qualitative explanation for the saturation of the peak growth observed in the experiment, indicating the importance of saturation effects at intermediate modulation times, and further highlight the advantage of momentum-resolved measurements for revealing parametric instabilities (Fig.~\ref{Fig_4}).

\subsection{Measured instability rates}
\begin{figure}
	\includegraphics{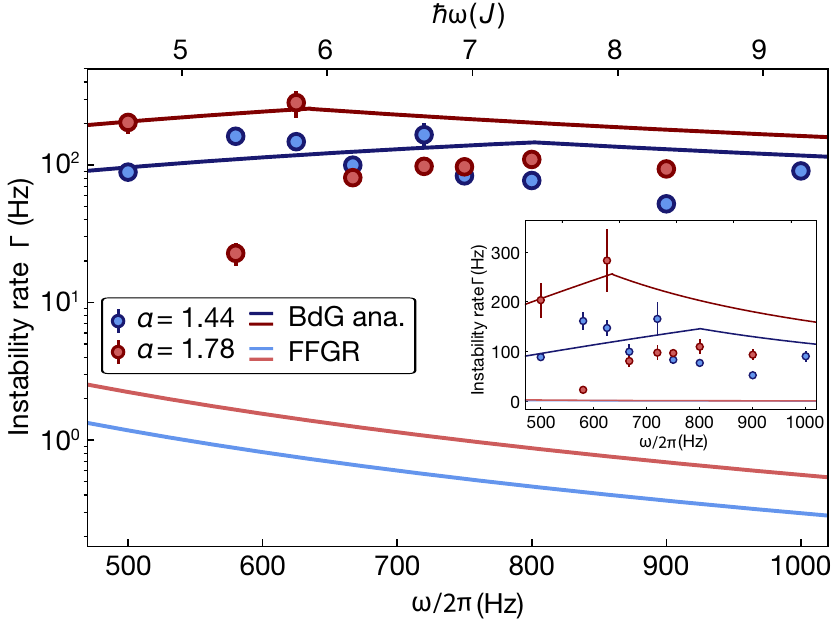}
	\vspace{-0.cm} \caption{Instability rates extracted from linear fits to the logarithmic peak amplitudes as a function of modulation time for $\alpha=1.44$ and $\alpha=1.78$. The dark solid lines are the theoretical rates calculated from Eqs~(\ref{eq:gamma_BdG1})-(\ref{eq:gamma_BdG2}) for $g=11.5 J$ and the bright solid lines are a FFGR calculation for the same modulation parameters according to Eq.~(\ref{gamma_FFGR}). The inset shows the same data on a linear scale to reveal the frequency dependence of the measured rates. Error bars indicate the standard deviation.}
	\label{Fig_10}
\end{figure}

Motivated by the numerical analysis discussed above, we quantitatively study the exponential growth rate of the most unstable mode at short times, before saturation effects dominate the dynamics. To determine the saturation point of the amplitudes and the instability rates from the experimental data, we fit a piecewise function consisting of a linear and a constant part to to the logarithm of the peak amplitude $A^x$~\cite{supplements}.
In BdG theory, the peak amplitudes along the $x$-lattice direction grow in time according to $A^x \propto e^{2\Gamma t}$. 
Hence, from the fitted slopes $m$ (in the regime $t<t_s$), we extract the instability rates as $\Gamma = m/2$ and average over the left and right peak. In Fig.~\ref{Fig_10}, we show a comparison between the experimentally observed rates and the theoretical expectation based on the analytical BdG predictions [Eqs~(\ref{eq:gamma_BdG1})-(\ref{eq:gamma_BdG2})]. We further display the long-time heating rates, which have shown to be captured by Floquet Fermi's Golden Rule (FFGR) in Ref.~\cite{Reitter:2017ij}. For completeness the relevant equations are summarized in the subsequent section.

We find that the experimental rates [dots in Fig.~\ref{Fig_10}] are of similar magnitude as compared to the BdG predictions and about several orders of magnitude larger than the FFGR predictions~\cite{Bilitewski:2015kx}. The inset of Fig.~\ref{Fig_10} shows all rates on a linear scale, to illustrate the dependence on the modulation parameters. From BdG theory, it is expected that the rates have a maximum at the saturation frequency $\omega_\mathrm{sat}$ [where $\hbar \omega$ equals the effective bandwidth]. This maximum value is expected to increase with larger modulation amplitude. 

Although the measured rates lie in the same order of magnitude as the BdG rates, the BdG-predicted parameter dependence is not directly confirmed by the measurements. Given the narrow window determined by $t_s$ during which the most unstable mode dominates, this is not surprising. More importantly,  the discrepancy between the measured rates and the FFGR predictions shows that we are indeed probing the system on sufficiently short timescales, where instabilities are mostly driven by coherent processes.

\subsection{Floquet Fermi's Golden Rule (FFGR)}

Heating on long timescales dominated by incoherent processes is well described by FFGR as studied in detail in Ref.~\cite{Reitter:2017ij}. The exponential decay of the BEC atoms caused by the modulated lattice can be described as

 \[
\dot{N}(t)=- \Gamma_{\text{FFGR}}N(t),
\]
\noindent where $ \Gamma_{\text{FFGR}}=\kappa N^{2/5}$. The loss rate $\kappa$ is independent of the atom number and given by:

\begin{equation}
\kappa(\alpha, \omega)=\frac{128}{105} \zeta^{-3/5} \left(\frac{15 \pi a_s}{8 d}\right)^{7/5} \left(\frac{\hbar \bar{\omega}}{E_R}\right)^{6/5} \frac{E_R}{\hbar} \sum_l\gamma_l
\label{gamma_FFGR}
\end{equation}

\noindent with
\begin{align}
\zeta &= d \int |w_0(x)|^4~dx, \nonumber\\
\bar{\omega} & = \sqrt[3]{\omega_r^2\omega_z}, \nonumber\\
\gamma_l &= 6 u_l g(s_l) \left( \frac{\zeta J \mathcal{J}_l(\alpha)}{l \hbar \omega}\right)^2, \nonumber\\
g(s_l) &= \left(\frac{1}{2} - \frac{1}{\pi}\arcsin(1-2s_l)\right)-\frac{2 s_l + 6}{3\pi}\sqrt{s_l(1-s_l)}, \nonumber\\
s_l &= \frac{l \hbar \omega}{8 J \mathcal{J}_0(\alpha)}, \nonumber\\
u_l &= \left\{ 
\begin{matrix}  0.75 & \mathrm{for} & l ~ \mathrm{even} 
\nonumber\\
 0.15 & \mathrm{for} & l ~ \mathrm{odd}
\end{matrix}, 
\right\},
\end{align}

\noindent $w_0(x)$ denoting the wannier function of the lowest band and $\mathcal{J}_\nu$ being the $\nu$-th Bessel function of the first kind describing $\nu$-photon scattering processes. For the model parameters used in our study, the rates are dominated by processes up to second order, keeping contributions up to $l=2$. The FFGR rates are displayed in Fig.~\ref{Fig_10} [bright solid lines]. It is evident that the FFGR rates are several orders of magnitude smaller than the BdG rates [dark solid lines]. This failure of FFGR to capture the short-time heating dynamics can be traced back to the exponentially dominating short-time parametric instability effect which, in contrast to FFGR, is sustained by coherent dynamics. We note in passing that such parametric instabilities can only occur in bosonic systems [or, more precisely, in arbitrary systems with bosonic elementary excitations].

\end{appendix}

\onecolumngrid\newpage

\renewcommand{\thefigure}{S\arabic{figure}}
 \setcounter{figure}{0}
\renewcommand{\theequation}{S.\arabic{equation}}
 \setcounter{equation}{0}
 \renewcommand{\thesection}{S\arabic{section}}
\setcounter{section}{0}

\onecolumngrid
\clearpage
\begin{center}
\noindent\textbf{Supplementary Material for:}
\\\bigskip
\noindent\textbf{\large{%
       Parametric instabilities of interacting bosons in periodically-driven 1D optical lattices}}
\\\bigskip

K. Wintersperger$^{1,2,*}$, M. Bukov$^{3,*}$, J. N\"ager$^{1,2}$, S. Lellouch$^{4,5}$, \\ E. Demler$^{6}$, U. Schneider$^{7}$, I. Bloch$^{1,2,8}$, N. Goldman$^{4}$, M. Aidelsburger$^{1,2}$\\\vspace{0.3em}

\small{$^{1}$\,Fakult\"at f\"ur Physik, Ludwig-Maximilians-Universit\"at M\"unchen, Schellingstra{\ss}e 4, 80799 M\"unchen, Germany}\\
\small{$^{2}$\,\emph{Munich Center for Quantum Science and Technology (MCQST), Schellingstr. 4, 80799 M\"unchen, Germany}}\\
\small{$^{3}$\,\emph{Department of Physics, University of California, Berkeley, CA 94720, USA}}\\
\small{$^{4}$\,\emph{Center for Nonlinear Phenomena and Complex Systems,\\ Universit\'e Libre de Bruxelles, CP 231, Campus Plaine, 1050 Brussels, Belgium}}\\
\small{$^{5}$\,\emph{Laboratoire de Physique des Lasers, Atomes et Mol\'ecules,\\ Universit\'e Lille 1 Sciences et Technologies, CNRS, 59655 Villeneuve d'Ascq Cedex, France}}\\
\small{$^{6}$\,\emph{Department of Physics, Harvard University, Cambridge, MA 02138, USA}}\\
\small{$^{7}$\,\emph{Cavendish Laboratory, University of Cambridge, J.~J.~Thomson Avenue, Cambridge CB3 0HE, UK}}\\
\small{$^{8}$\,\emph{Max-Planck-Institut f\"ur Quantenoptik, Hans-Kopfermann-Stra{\ss}e 1, 85748 Garching, Germany}}\\
\small{$^{*}$\,these authors contributed equally to this work}
\end{center}
\bigskip
\bigskip
\twocolumngrid


\section{\label{sec:data_ana}Details of the data analysis and calibrations}

\subsection{\label{subsec:aborb_im}Analysis of the absorption images}

As described in the main text, the excitation peaks are evaluated from difference images (Fig.~\ref{Fig_2}b), which are obtained by subtracting the mean condensate profile at $t=0$ from each single image at time $t>0$ for a certain set of modulation parameters. 
The mean image at $t = 0$ is obtained as described below. Due to the high density of the Bose-Einstein condensate (BEC) in the center, the imaging light is completely absorbed there, leading to a saturation of the pixel values, especially at $t=0$, when the number of atoms in the BEC is maximal. 

In order to minimize statistical fluctuations in the mean position of the atom cloud, we shift the pixels such that the center of the BEC peak is at $(x,y)=(0,0)$. The center is extracted by fitting a two-dimensional (2D) Gaussian function to the BEC peak, which is truncated at some value $I_0$ to account for the saturation in the center: 
\begin{equation}
g(x,y) =\begin{cases} A e^{-\left(\frac{(x-x_0)^2}{2\sigma_x^2}+\frac{(y-y_0)^2}{2\sigma_y^2}\right)}+y_0 & g(x,y)< I_0 \\
I_0 & g(x,y) \ge I_0 
\end{cases} \nonumber
\end{equation}

For each parameter set, we determine a saturation value $I_0$ from all images at $t=0$ and apply it to all images in this data set. We start by fitting a truncated Gaussian to each image at $t=0$ separately and take the mean over all $I_0^j$ as a starting value for the following fit. Then, we fit a truncated Gaussian to all images simultaneously with the constraint that the saturation value $I_0$ has to be the same in all images, whereas all other parameters can vary. This procedure results in the final value $I_0^{*}$. To determine the center positions, we fit a truncated Gaussian $g^{*}(x,y)$ with saturation value $I_0^{*}$ to all images and use these to shift the pixels in each image such that all center positions are at coordinates $(0,0)$ and truncate the pixel counts on all images at $I_0^{*}$. The centered and truncated images are then averaged to give a mean $t=0$ image for each parameter set.

For the subtraction, we take each single image at every $t>0$ and first fit the previously defined truncated Gaussian $g^{*}(x,y)$ with the fixed $I_0^{*}$ to determine the center of the BEC peak. Then, the profiles are again shifted to move their center to $(0,0)$ and the pixel values are truncated at $I_0^{*}$. From the centered, truncated image we then subtract the mean $t=0$ image in order to obtain a difference image as shown in Fig.~\ref{Fig_2}b in the main text. Due to the centering of the images the statistical fluctuations due to drifts of the BEC position are suppressed. Hence, fluctuations of the position of the excitation peaks that are detected for $t>0$ correspond to the uncertainty of this peak position relative to the BEC position and is reflected in the standard error of the corresponding mean value at this time. 

The conversion from pixel values to quasimomentum is achieved by taking an absorption image after switching of the lattice abruptly and a time-of-flight (TOF) of $6\,$ms. This results in interference peaks, which are separated by $2\pi/d$. We determine the positions of these by Gaussian fits and average over several experimental realizations, which gives the width of the first Brillouin zone (BZ) in pixels. From the positions of the Bragg peaks, also the exact direction of the lattice on the camera can be determined. In Fig.~\ref{Fig_2}, the images were rotated by $6^{\circ}$ to match the lattice axis with the horizontal direction for better visualisation. In the data analysis, we sum up the pixels along an axis (or perpendicular to) that is turned by $6^{\circ}$ instead of rotating the images, which would lead to false empty pixels. 

As discussed in the main text, we expect that for modulation frequencies above the saturation frequency $\omega_{\text{sat}}$, the excitation peaks in momentum space should be located at $\pi/d$ with a tail towards smaller quasimomenta, resulting from excitations to other modes and from the width of the condensate at $t=0$, which has a finite (Gaussian) width of $\sigma_0\approx 0.2 \pi/d$. As shown in Fig.~\ref{Fig_4}, the measured saturation value of the most unstable mode along the lattice axis $q^x_{\text{mum}}$ is slightly smaller than $\pi/d$. This is most likely due to the short TOF used in the experiments, i.e., we measure a convolution of the momentum profile with the insitu density distribution of the BEC. This leads to a slight shift of the peak maxima in the TOF images towards the center of the BZ.

\subsection{\label{subsec:peak_ana}Details of the Peak analysis}

To analyze the peaks along the lattice direction, we integrate all difference images (Fig.~\ref{Fig_2}b) along the direction perpendicular to the lattice to obtain a 1D profile, as described in the main text. To extract the peak amplitude $A^x$ and peak position we fit an asymmetric Loretzian function 

\begin{align}
\label{eq:asymLorentz}
L(q^x)=\begin{cases}\tilde{A}\frac{w_1^2}{w_1^2+4(q^x_{\text{max}}-q^x)^2}+A_0 & q^x < q^x_{max} \\
\\[1pt]
\tilde{A} \frac{w_2^2}{w_2^2+4(q^x_{\text{max}}-q^x)^2}+A_0 & q^x \ge q^x_{max} \\ \end{cases}
\end{align}

\noindent where we define the peak amplitude as $A^x=\tilde{A}-|A_0|$. The asymmetry of the profiles arises from the thermal background inside the first BZ which effectively increases the profile height at the inner side of the peaks. To determine the transverse width $\Delta q^y$ of the peaks, we integrate the difference profile along the lattice direction inside a region of interest that contains only the left or right peak (excluding the negative pixels in the center) and fit a symmetric Lorentzian to each of the peaks: 
\begin{align}
L(q^y)=\tilde{B}\frac{(\Delta q^y)^2}{(\Delta q^y)^2+4(q^y_{\text{max}}-q^y)^2}+B_0.
\end{align} 

The end of the short-time regime $t_s$ (Fig.~\ref{Fig_3}) is obtained by fitting the following piecewise function to the logarithm of the peak amplitudes $\text{ln}(A^x)$ for the left and right peak independently:

\begin{align}
K(t) =\begin{cases} m t + b & t < t_s \\
m t_s + b & t \ge t_s \\
\end{cases}
\label{eq_SI_kink}
\end{align}

The final value for $t_s$ is defined as the average over the fit results for the left and right peak (rounded to integer multiples of the modulation period).

\begin{figure}[t]
\includegraphics{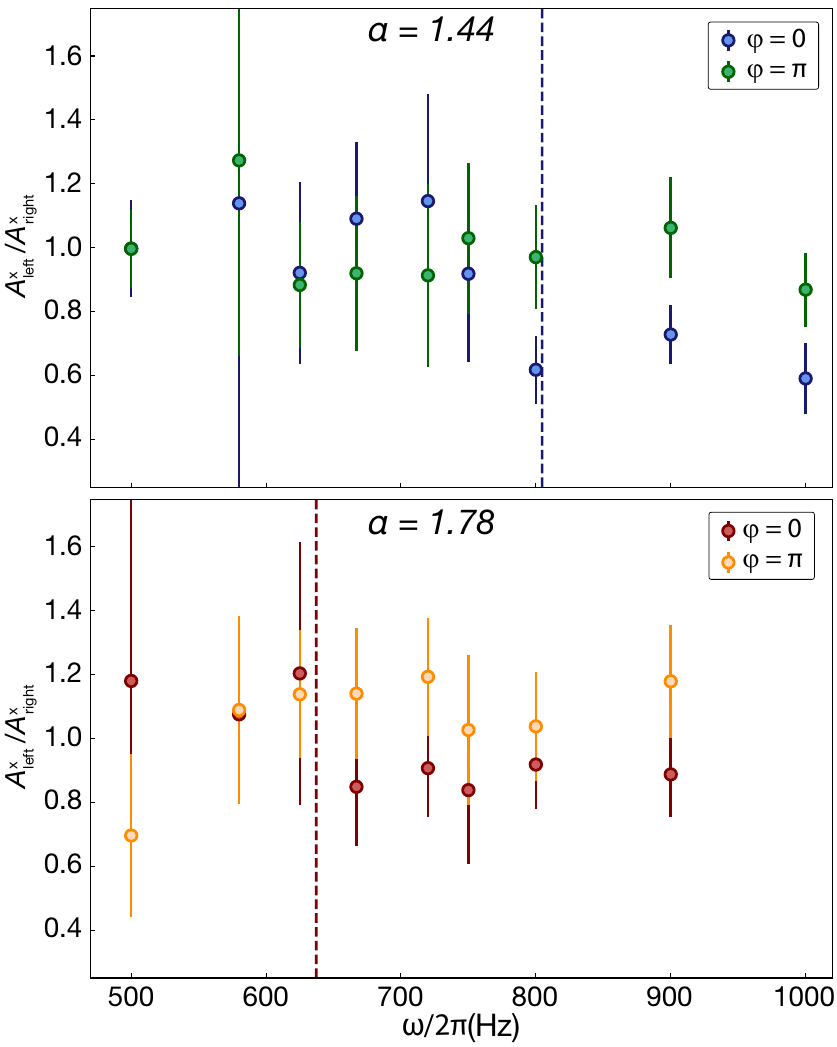}
\vspace{-0.cm} \caption{Ratio of the fitted amplitudes for the left and right excitation peak $A^x_{\text{left}}/A^x_{\text{right}}$ for two different phases $\varphi=0$ and $\varphi=\pi$ of the periodic modulation as a function of modulation frequency for $\alpha=1.44$ and $\alpha=1.78$. For each time step we average $\lessim\,5$ individual experimental realizations. The data points show the mean of all obtained asymmetry values $A^x_{\text{left}}/A^x_{\text{right}}$ for $t\leq t_s$. The error bars show the standard error of the weighted mean. The experimental results shown in the manuscript are averaged over $\varphi=0$ and $\varphi=\pi$. The dashed lines mark the saturation point, where $q^x_{\text{mum}}$ reaches the edge of the BZ, as in Fig.~\ref{Fig_4}. \label{Fig_asymmetry}}
\end{figure}

We find that for parameters, where $q^x_{\text{mum}}$ saturates at the band edge, $q^x\approx\pi/d$, the observed momentum profiles are slightly asymmetric, i.e. either the left or right excitation peak has a larger amplitude (Fig.~\ref{Fig_asymmetry}). We further observed that the asymmetry depends on the initial phase of the drive $\varphi$. It is known that for momentum components at the band edge any residual gradients or finite ramp times during the switching off of the lattices can result in an asymmetry in the bandmapped images. In order to reduce this systematic deviations we average over individual experimental realizations with $\varphi=0$ and $\varphi=\pi$. The final results show no residual systematic asymmetry (Figs~\ref{Fig_S1}-\ref{Fig_S3}).

\subsection{\label{subsec:atom_calib}Atom number calibration}

To determine the atom number, we measure the trapping frequencies in the absence of the lattice $[\tilde{\omega}_r/(2\pi)=23.2(3)\,\text{Hz},\,\tilde{\omega}_z/(2\pi)=189(3)\,\text{Hz}]$ and the Thomas-Fermi radii as a function of the scattering length $a_s$ and fit the number of atoms according to the Thomas-Fermi prediction 

\begin{equation}
R^{i}_{TF} = \left( \frac{15 N a_s \hbar^2 \omega_x \omega_y \omega_z}{m^2 \omega_i^5}\right)^{1/5}.
\end{equation}
We obtain $N = 3.7(4) \times 10^{5}$, where the error bar mainly stems from atom number fluctuations. Note, that for the atom number calibration, the harmonic confinement is reduced compared to the measurements with the lattice that are presented in the main text. The trapping frequencies are determined by monitoring the center of mass motion of the BEC in-situ after a displacement in the trap. This gives the harmonic trapping frequencies along the axes of the dipole trap beams. In the horizontal plane the trap is rather symmetric with $\tilde{\omega}_x/(2\pi)=24.0(5)\,\text{Hz}$ and $\tilde{\omega}_y/(2\pi)=22.3(3)\,\text{Hz}$, so we consider their mean value $\tilde{\omega}_r$. The same is true for the situation with the lattice being on.

\section{\label{sec:modulation_time}Peak parameters vs.~modulation time}

\begin{figure*}[h!]
\includegraphics[width=0.8\textwidth]{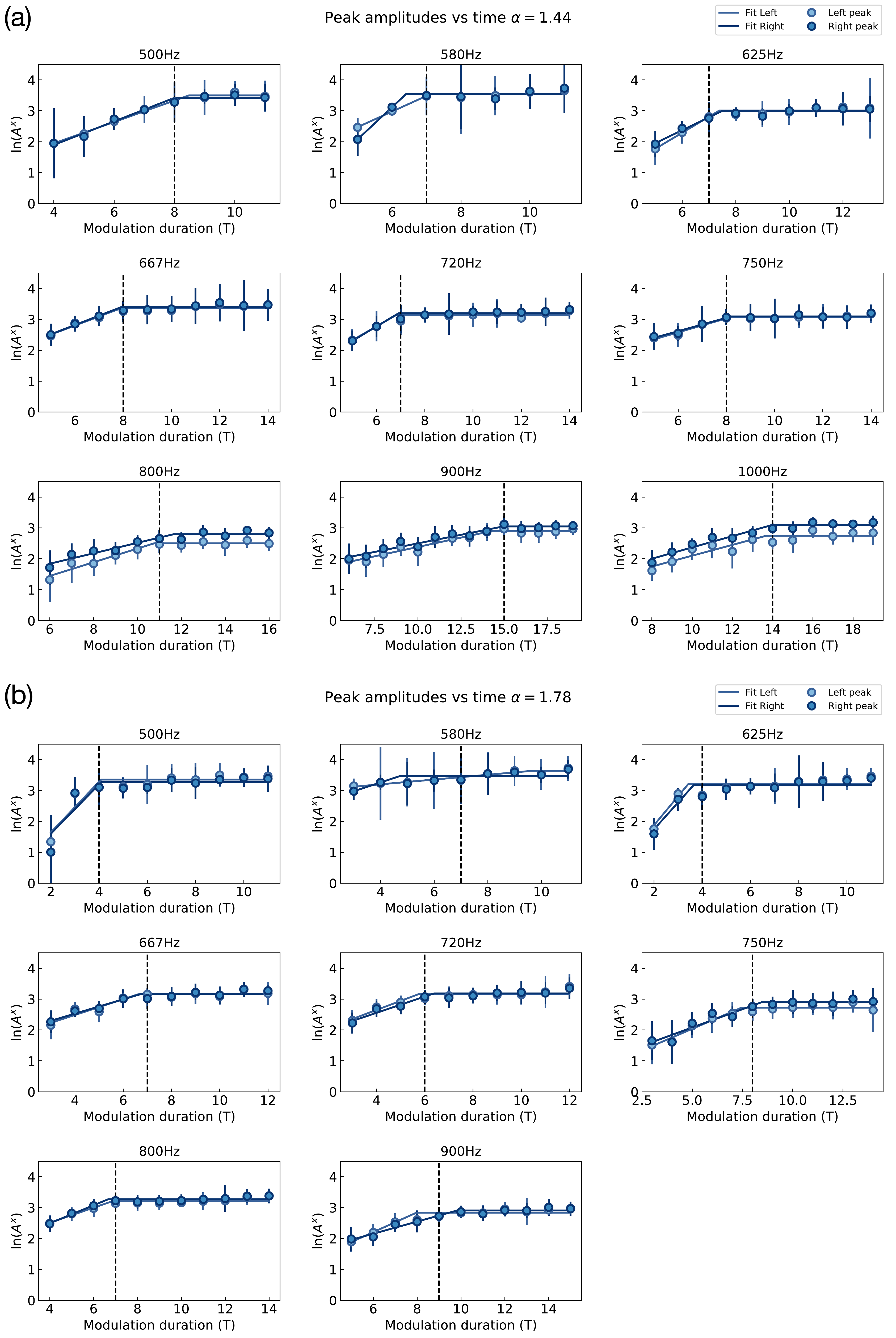}
\vspace{-0.cm} \caption{Logarithmic plot of the peak amplitudes $A^x$ for $\alpha=1.44$ (a) and $\alpha=1.78$ (b) versus modulation time. Each point is an average over $\sim10$ individual experimental realizations, $\lessim\,5$ realizations for a modulation phase $\varphi=0$ and $\lessim\,5$ with $\varphi=\pi$. The error bars indicate the standard error. Solid lines are fits to the data according to Eq.~(\ref{eq_SI_kink}), the black dashed line marks the mean kink value $t_s$ rounded to integer values. \label{Fig_S1}}
\end{figure*}

\begin{figure*}[h!]
\includegraphics[width=0.8\textwidth]{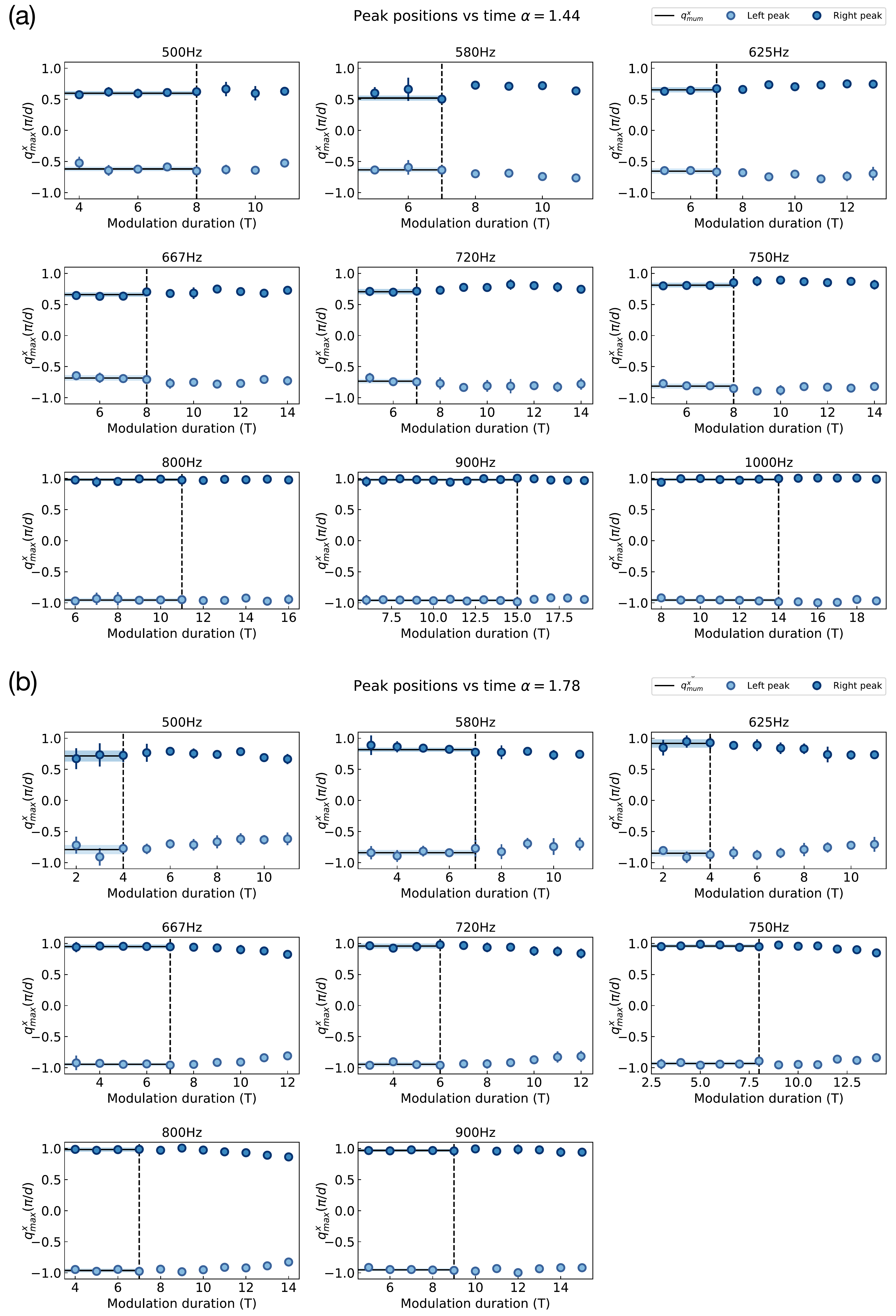}
\vspace{-0.cm} \caption{Peak positions $q^x_{\text{max}}$ for $\alpha=1.44$ (a) and $\alpha=1.78$ (b) as a function of modulation time. Each point is an average over $\sim10$ individual realizations, $\lessim\,5$ realizations for a modulation phase $\varphi=0$ and $\lessim\,5$ with $\varphi=\pi$. The error bars indicate the standard error. The black solid lines denote $q^x_{\text{mum}}$ the weighted mean over the points up to the transition time $t_s$ (vertical black dashed line), the shaded bars indicate the standard error of the weighted mean.\label{Fig_S2}}
\end{figure*}

\begin{figure*}[h!]
\includegraphics[width=0.8\textwidth]{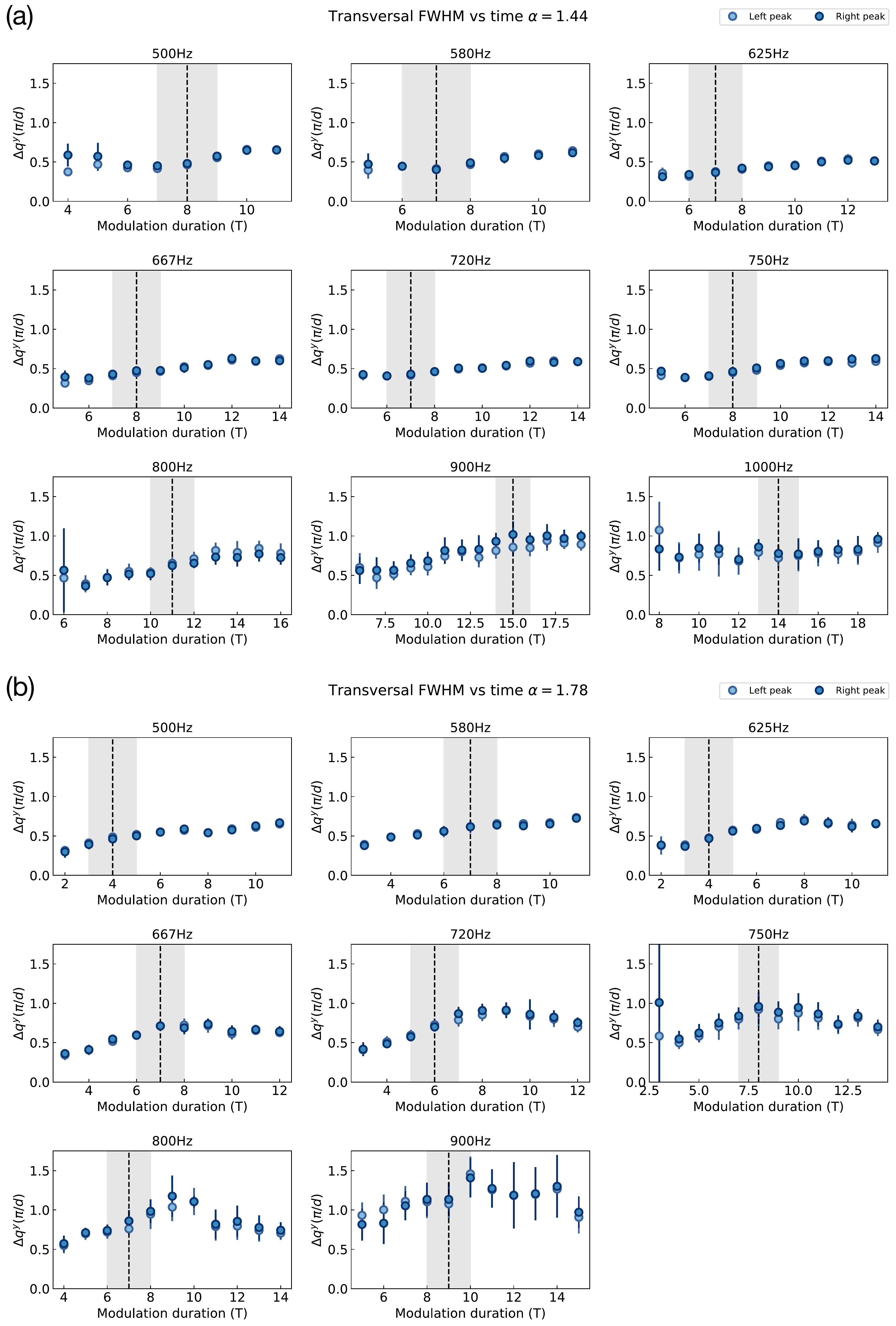}
\vspace{-0.cm} \caption{Transverse full-width at half maximum (FWHM) versus modulation time. Each point is an average over $\sim10$ individual experimental realizations, $\lessim\,5$ realizations for a modulation phase $\varphi=0$ and $\lessim\,5$ with $\varphi=\pi$. The error bars indicate the standard error. The black, dashed line marks the end of the short-time regime $t_s$ and the grey shaded area extends from $t_s - 1$ to $t_s + 1$. The points in this range were averaged to obtain the data points in Fig.~\ref{Fig_4} (lower panel). \label{Fig_S3}}
\end{figure*}

In Fig.~\ref{Fig_S1} the logarithm of the fitted peak amplitudes versus modulation time is shown for all modulation frequencies and modulation amplitudes $\alpha=1.44$ and $\alpha=1.78$ together with the fitted piecewise functions [defined in Eq.~(\ref{eq_SI_kink})]. The dashed lines mark again the transition time $t_s$ where the short time regime ends, which is the mean value of the fitted kink positions for the left and right peak rounded to integer multiples of the modulation period. In general, the peak amplitudes decrease with the modulation frequency, also the errorbars become smaller. We attribute this to the enhanced influence of incoherent processes at small modulation frequencies, which are closer to the bandwidth of the static 1D lattice ($4J/h = 0.43\,\text{kHz}$).
The fitted peak positions $q^x_{\text{max}}$ are shown in Fig.~\ref{Fig_S2} where the black dashed lines again mark the end of the short-time regime, $t_s$. The black lines denote $q^x_{\text{mum}}$ which is the mean over all positions up to $t_s$. The data points in the upper panel of Fig.~\ref{Fig_4} are the average over $q^x_{\text{mum}}$ for the left and right peak for each set of modulation parameters.
In Fig.~\ref{Fig_S3} we plot the transverse width $\Delta q^y$ versus modulation time, the grey shaded area denotes the data range, which is averaged to obtain the data points in the lower panel of Fig.~\ref{Fig_4}. In general, the transverse width is nearly constant for frequencies below the saturation frequency, as discussed in the main text. For frequencies larger than $750\,\text{Hz}$ for $\alpha=1.44$ and $625\,\text{Hz}$ for $\alpha=1.78$, the peak widths increase but they also fluctuate more as indicated by the larger error bars.


\clearpage

\section{\label{sec:dynamics_theory} Theoretical methods and analytical treatment of instabilities in the periodically-driven Bose-Hubbard model}

In this section, we give details about the theoretical methods and results used in the main text to discuss the dynamics and stability properties of the periodically-driven weakly-interacting Bose-Hubbard model (BHM). For simplicity we set $\hbar=1$, the lattice constant $d=1$ and $ d_{\perp}=1$ in the entire section, cf.~Sec.~\ref{sec:constants}.

The BHM is defined by the Hamiltonian
\begin{align}
\hat{H}(t) = &-J \sum_{\langle ij\rangle,\rp} (\hat{a}^{\dagger}_{i,\rp} \hat{a}_{j,\rp} + \mathrm{h.c.}) +  \hat{H}_\perp \nonumber\\
& + K \cos (\omega t + \varphi) \sum_{j,\rp} j\hat{n}_{j,\rp} \nonumber\\
& + \frac{U}{2} \sum_{j,\rp} \hat{n}_{j,\rp} (\hat{n}_{j,\rp} -1),
\label{eq:H(t)}
\end{align}
where $\ah_{i,\rp}$ annihilates a particle at lattice site $i$ and transverse position $\rp = l_\perp d_{\perp} \hat{\mathbf{e}}_\perp$. 
$J>0$ denotes the tunneling amplitude of nearest-neighbor hopping along the $x$ direction, $\hat{H}_\perp=\sum_{j,{\bf q}^\perp}\frac{ \left({\bf q}^\perp\right)^2}{2m} \hat{n}_{j,{\bf q}^\perp}$ describes a free-particle kinetic part along transverse directions, and $U>0$ is the repulsive on-site interaction strength. The time-periodic modulation has amplitude $K$, phase $\varphi$, and frequency $\omega=2\pi/T$ with $T$ the driving period. 

Transforming the system to the rotating frame~\cite{Lellouch:2017its}, we eliminate the external shaking at the expense of introducing a periodically-modulated hopping term
\begin{align}
\hat{H}_\mathrm{rot}(t) = &-J \sum_{\langle ij\rangle,\rp} (\mathrm e^{-i\alpha\sin(\omega t + \varphi)}\hat{a}^{\dagger}_{i,\rp} \hat{a}_{j,\rp} + \mathrm{h.c.}) \nonumber\\
& + \hat{H}_\perp + \frac{U}{2} \sum_{j,\rp} \hat{n}_{j,\rp} (\hat{n}_{j,\rp} -1),
\label{eq:H_rot(t)}
\end{align} 
with the dimensionless coupling strength $\alpha=K/\omega$.

In momentum space, the free-system [$U\!=\!0$] is governed by the time-dependent dispersion
\[
\varepsilon_{\qbf}(t)=-2J\cos \left(q^x - \alpha\sin(\omega t + \varphi) \right) + \frac{\left({\bf q}^\perp\right)^2}{2m},
\] 
oscillating at the driving frequency $\omega$.

Our study relies on three different methods: 
\begin{enumerate}
	\item[A.] time-dependent Bogoliubov-de Gennes (BdG) theory [Sec.~\ref{subsec:Bog_lin}],
	\item[B.] the Weak Coupling Conserving Approximation (WCCA) [Sec.~\ref{subsec:WCCA}],
	\item[C.] the Truncated Wigner Approximation (TWA) [Sec.~\ref{subsec:TWA}].
\end{enumerate}

\subsection{\label{subsec:Bog_lin} Time-Dependent Bogoliubov-de Gennes Theory (BdG)}

In the weakly-interacting regime we apply Bogoliubov theory. Weakly-interacting bosons at ultracold temperatures form a Bose-Einstein condensate (BEC) which is macroscopically occupied, whereas the population of all other modes remains small compared to that of the condensate. This scale separation motivates the Bogoliubov approximation:
\begin{equation}
\label{eq:Bog_approx}
\hat a_{\qbf_\mathrm{BEC}} = \hat b_{\qbf_\mathrm{BEC}} + \sqrt{N_\mathrm{BEC}(t)},\qquad \hat a_{\qbf\neq \qbf_\mathrm{BEC}} = \hat b_{\qbf\neq \qbf_\mathrm{BEC}},
\end{equation}
where $N_\mathrm{BEC}(t)$ denotes the BEC population, $\qbf_\mathrm{BEC}$ is the mode in which condensation occurs, and $\hat b^\dagger_\qbf$ creates a particle of momentum $\qbf$ on top of the BEC background.

Plugging this ansatz into the Hamiltonian~\eqref{eq:H_rot(t)}, and keeping terms to quadratic order in the operators $\hat b_\qbf$, we arrive at the time-dependent Bogoliubov Hamiltonian 
\begin{equation}
\label{eq:H_Bog(t)}
H_\mathrm{B}(t) = \sum_\qbf (\varepsilon_\qbf(t)-\varepsilon_{\qbf_\mathrm{BEC}}(t) + g)\hat b^\dagger_\qbf\hat b_\qbf + \frac{g}{2}\sum_\qbf \hat b^\dagger_\qbf \hat b^\dagger_{-\qbf} + \mathrm{h.c.}
\end{equation}
Here $g\!=\!U n$, $n\!=\!N_\mathrm{BEC}(t\!=\! 0)/V$ with $V$ the volume of the system.
The time-dependent shift $\varepsilon_{\qbf_\mathrm{BEC}}(t)$ arises from the chemical potential $\mu(t) = \varepsilon_{\qbf_\mathrm{BEC}}(t) + g$ which is fixed by the condition that the linear term (in $\hat b_\qbf,\hat b^\dagger_\qbf$) in the Bogoliubov expansion vanishes. The same result can be obtained from a linearization of the time-dependent Gross-Pitaevskii equation around the BEC wavefunction~\cite{Lellouch:2017its,Creffield:2009cps}.

In the Bogoliubov approximation, the dynamics of the system is governed by the time-dependent Bogoliubov mode functions $u_\qbf(t)$ and $v_\qbf(t)$, defined through
\begin{eqnarray*}
\hat b_\qbf(t) = u_\qbf(t) \hat \gamma_\qbf + v_{-\qbf}^*(t) \hat \gamma^\dagger_{-\qbf},\quad |u_\qbf(t)|^2-|v_\qbf(t)|^2 = 1.
\end{eqnarray*}
Using Heisenberg's equations of motion (EOM) $i\dot{\hat{b}}_\qbf(t)=[\hat b_\qbf(t),H_\mathrm{B}(t)]$, we arrive at the Bogoliubov-de Gennes (BdG) equations:
\begin{widetext}
\begin{equation}
i \partial_t \left( \begin{matrix} u_\qbf  \\ v_\qbf \end{matrix} \right)
=\left( 
\begin{matrix}  \varepsilon_\qbf(t)-\varepsilon_{\qbf_\mathrm{BEC}}(t)+g & g \\ 
-g & -\varepsilon_{-\qbf}(t)+\varepsilon_{\qbf_\mathrm{BEC}}(t)-g \end{matrix} 
\right)
\left( \begin{matrix} u_\qbf  \\ v_\qbf \end{matrix} \right).
\label{eq:BdGE_EOM}
\end{equation}
\end{widetext}
The BdG equations~\eqref{eq:BdGE_EOM} describe the quasi-particle dynamics, and include micromotion effects. 

Due to the time-periodicity of Eq.~\eqref{eq:BdGE_EOM}, Floquet theory applies. We can thus focus on the stroboscopic fundamental matrix $\Phi(T)$, which is obtained by time-evolving Eq.~(\ref{eq:BdGE_EOM}) over a single period $T$. We denote the eigenvalues of $\Phi(T)$ by $\epsilon_{\qbf}$. The appearance of eigenvalues with positive imaginary parts indicates a dynamical instability, i.e.~an exponential growth of the corresponding mode(s), characterized by the rate $\mathrm{Im}\; \epsilon_{\qbf}$.

We then define the instability rate as the maximum growth rate: 
\begin{equation}
	\Gamma\equiv \underset{{\qbf}}{\mathrm{max}}\ \mathrm{Im}\; \epsilon_{\qbf}.
	\label{eq:GammaDef}
\end{equation}
It is independent of the reference frame, and governs the stroboscopic dynamics of the mode functions $u_\qbf$ and $v_\qbf$. Notice that the number of exited atoms, $n(t)\!\sim\! \sum_{{\bf q}}\vert v_\qbf(t)\vert^2$, has an instability rate $2\Gamma$. 

Let us also introduce the most unstable mode
\begin{equation}
	\qbf_\mathrm{mum}\equiv \underset{\qbf}{\mathrm{argmax}}\ \mathrm{Im}\; \epsilon_{\qbf},
	\label{eq:qmumDef}
\end{equation}
which dominates the long-time BdG dynamics because all other modes grow more slowly. In the main text, $\qbf_\mathrm{mum}$ is shown to provide clear experimental signatures of parametric instabilities. 

Details on the explicit derivation of the instability rate and the maximally-unstable mode for the BHM can be found, e.g., in Ref~\cite{Lellouch:2017its}.

\subsection{\label{subsec:WCCA} Weak Coupling Conserving Approximation (WCCA)}

The Bogoliubov approximation holds under the condition that the condensate occupation remains large throughout the duration of interest. A prerequisite for this is the presence of weak enough interaction strength $U/J\lesssim 1$, which renders the occurrence of quasiparticle collisions rare. Indeed, the Bogoliubov approximation works well when the latter do not play an important role, and thus can be neglected. This is the case, for instance, for weakly interacting bosons in equilibrium.

Away from equilibrium, excitations may form induced by the drive, leading to fast depletion of the macroscopically occupied condensate, even though the interactions may still remain small (compared to the bare hopping) at all times. The condensate depletion is enhanced when resonant transitions between the condensate mode and higher-energy states occur; such transitions can be stimulated by a periodic drive. 

Conceptually, one of the major drawbacks of the time-dependent Bogoliubov approximation is the lack of particle number conservation~\cite{PartNum}. This means that, even though it may capture the onset of condensate depletion, the approximation is doomed to fail in correctly predicting the dynamics of the quasiparticle distribution during the later stages of the evolution. One reason for this lies in the fact that the BdG EOM assume a constant in time, and thus endless, reservoir of particles provided by the macroscopically occupied condensate mode.

An elegant way to restore particle number conservation is provided by the Weak Coupling Conserving Approximation (WCCA). It represents a minimal extension of the linearized Bogoliubov equations which conserves the corresponding global $U(1)$ symmetry at all times. The WCCA EOM is a set of coupled differential equations for the time-dependence of the condensate wavefunction $\phi(t)$ and the quasiparticle correlators $F_{11}$ and $F_{12}$, defined as
\begin{align}
\label{eq:WCCA_vars}
&\phi(t) =\langle \hat a_{\qbf_\mathrm{BEC}} \rangle, \\
&F_{11}(t;{\bf q}) = \frac{1}{2}\langle\{\hat a_{\bf q}(t),\hat a_{\bf q}^\dagger(t)\}\rangle_c=\frac{1}{2}\langle\{\hat b_{\bf q}(t),\hat b_{\bf q}^\dagger(t)\}\rangle, \notag \\
&F_{12}(t;{\bf q}) = \langle\{\hat a_{\bf q}(t),\hat a_{-{\bf q}}(t)\}\rangle_c= \langle\{\hat b_{\bf q}(t),\hat b_{-{\bf q}}(t)\}\rangle,\notag
\end{align}
where, $\hat{a}_{\bf q}$ describes the original bosonic degrees of freedom, while $\hat{b}_{\bf q}$ denotes the quasiparticle excitations [see Eq.~\eqref{eq:Bog_approx}]. The subscript $_c$ denotes the connected correlation function $\langle AB\rangle_c = \langle AB\rangle - \langle A\rangle \langle B\rangle$ in the Bogoliubov ground state [we work in the Heisenberg picture].

The WCCA equations can be derived by Legendre-transforming the action of the BHM on the Keldysh contour with respect to both the order parameter (condensate field) $\phi$, and the quasiparticle propagators $F_{11}$ and $F_{12}$~\cite{Babadi:2017dcs,Bukov:2015dvs}. The resulting effective action, which contains all two-particle irreducible diagrams, is then expanded to leading order in the interaction strength $U$. Minimizing this effective action w.r.t.~the condensate field and the propagators, one arrives at the following system of coupled integro-differential equations~\cite{Bukov:2015dvs}:
\begin{widetext}
	\begin{align}
	\label{eq:WCCA_EOM}
	i\partial_t\phi(t) \!=& [\varepsilon_{\qbf_\mathrm{BEC}}(t) - \mu(t)]\phi(t)
	+ \frac{U}{V}\bigg[ \left[\phi(t)\right]^*\left[\phi(t)\right]^2
	+ 2\phi(t) \sum_{{\bf q'}} F_{11}(t;{\bf q'}) + \left[\phi(t)\right]^*\sum_{{\bf q'}} F_{12}(t;{\bf q'}) \bigg],\\
	\partial_t F_{11}(t;{\bf {\bf q}}) \!=& 2\text{Im}\left\{  \frac{U}{V}\left( \left[\phi(t)\right]^2 + \sum_{{\bf q'}} F_{12}(t;{\bf q'})\right)\left[F_{12}(t;{\bf q})\right]^* \right\},\nonumber\\
	i\partial_tF_{12}(t;{\bf q}) \!=& \Bigg\{ [\varepsilon_{\bf q}(t)\!+\!\varepsilon_{-{\bf q}}(t)\!-\!2\mu(t)\!]F_{12}(t;{\bf q})
	\nonumber\\
	&\!+\! \left. 2\frac{U}{V}\left[ 2\left(\! |\phi(t)|^2 \!+\! \sum_{{\bf q'}} F_{11}(t;{\bf q'}) \right)\!F_{12}(t;{\bf q})		
	\!+\! \left(\! \left[\phi(t)\right]^2 \!+\! \sum_{{\bf q'}} F_{12}(t;{\bf q'})\right)\! F_{11}(t;{\bf q}) \right] \right\},\nonumber
 	\end{align}
\end{widetext}
where we denoted by $^*$ the complex conjugation. 

The initial condition is given by the Bogoliubov ground state, together with $|\phi(t\!=\!0)|^2=N_\mathrm{BEC}(0)$. Notice that the presence of the chemical potential in the WCCA EOM is irrelevant to any observables, due to the $U(1)$-symmetry being conserved at all times, in contrast to the BdG EOM, for which $\mu(t)$ is crucial to recover the correct spectrum of excitations. Hence, for the sake of a proper comparison with the BdG approximation, we may set $\mu(t) = \varepsilon_{\qbf_\mathrm{BEC}}(t) + g$, which also fixes the reference frame.

One can convince oneself that, if one neglects all terms involving ${\bf q}'$-summations in Eqs.~\eqref{eq:WCCA_EOM}, the equation for the condensate wavefunction $\phi(t)$ decouples from the other two. Moreover, one may further recognize it as the Gross-Pitaevskii equation in the presence of the periodic drive. Using the definition~\eqref{eq:WCCA_vars}, it follows that $F_{11}(t;{\bf q})=1/2(|u_{\bf q}(t)|^2+|v_{\bf q}(t)|^2)$ and $F_{12}(t;{\bf q})=u_{\bf q}(t)v_{\bf q}(t)$, and hence the remaining equations for $F_{11}$ and $F_{12}$ are equivalent to the BdG EOM from Eq.~\eqref{eq:BdGE_EOM}. Going back to the complete equations, we see that the ${\bf q}'$-summations represent the essential coupling between the condensate and the quasiparticle excitations, necessary to restore particle number conservation to leading order in $U$. The conserved quantity itself is the total number of condensed and excited atoms
\begin{align}
N &= |\phi(t)|^2 + \sum_{\bf q} n_{\bf q}(t) \notag \\
&= |\phi(t)|^2 + \sum_{\bf q} \left( F_{11}(t;{\bf q}) - \frac{1}{2}\right) = \mathrm{const.}
\end{align}

Despite the advantages of the WCCA EOM over the BdG equations discussed above, let us make some remarks about the applicability and usefulness of the WCCA. First, the WCCA EOM are not amenable to simple analytical treatment, due to their non-locality in momentum space. Second, unless condensate depletion is suppressed, e.g.~by the existence of a pre-thermal phase~\cite{Bukov:2015dvs}, the WCCA is also expected to be valid only in the short-time limit, since it misses the collisions between quasiparticles which appear first at order $\mathcal{O}(U^2)$. Thus, within the WCCA, a periodically driven system cannot thermalize which points out that the WCCA does not capture the later stages of the heating process. Going to order $U^2$ has been done recently in the large-$N$ limit of the periodically-driven $O(N)$ model~\cite{Chandran:2016s,Weidinger:2017ffs}, and the existence of long-lived pre-thermal plateaus has been shown in the limit where the drive frequency is the largest energy scale in the problem.    

\subsection{\label{subsec:TWA} Truncated Wigner Approximation (TWA)}

Since the exact quantum equations of motion for the BHM are often too hard to analyze, it may be advantageous to perform a semi-classical analysis. This proves particularly useful when the initial condition is deep into the superfluid phase, as the latter is well captured by the GPE, a classical nonlinear wave equation. Recently, it has been demonstrated that thermalization in classical Floquet systems behaves very similarly to their quantum counterparts~\cite{Howell:2018uys,Notarnicola:2018fts,Rajak:2018tas}.

An systematic way of deriving the leading-order quantum corrections in the semiclassical limit, is to apply the Truncated Wigner Approximation (TWA). This method has the advantage that it is both particle-number conserving [since the GPE conserves $U(1)$ symmetry] and, at the same time, it is capable of capturing the thermalizing dynamics induced by the periodic modulation. Since there is extensive literature on the TWA already, below we briefly summarize the main ideas useful to our analysis, and refer the interested reader to Refs.~\cite{Blakie:2008iss,polkovnikov2010phases}.

From translational invariance, it follows that the lowest-energy state of the BHM must necessarily be uniform in real space and, therefore, in momentum space the entire weight is carried by the $\qbf_\mathrm{BEC}=0$ mode. Thus, in the classical limit, the uniform condensate field obeys the following time-dependent GPE:
\begin{equation}
i\partial_t\phi(t) = (\varepsilon_{\bf q=0}(t)-\mu(t))\phi(t) + \frac{U}{V}|\phi(t)|^2\phi(t).
\end{equation}
Similar to the WCCA EOM, the presence of the chemical potential in the GPE is irrelevant to any observables, due to the $U(1)$-symmetry of the GPE. For instance $|\phi(t)|^2=N_\mathrm{BEC}\approx N$ at all times. 

In quantum mechanics, however, a small but finite fraction of atoms always occupies the excited modes of the system, even for weak interactions, known as quantum depletion. Let us again denote the Bogoliubov mode functions by $u_{\bf q}(t)$ and $v_{\bf q}(t)$, and recall that the complete atomic field expansion reads
\begin{eqnarray}
\hat a_j &=& \frac{1}{\sqrt{V}}\sum_{\bf q} \hat a_{\bf q}\mathrm e^{-i {\bf q}\cdot {\bf r}_j }\nonumber \\
&=& \frac{1}{\sqrt{V}} \left(\sqrt{N_{{\bf q}=0}}+\hat b_{{\bf q}=0}\right) + \frac{1}{\sqrt{V}}\sum_{\bf q\neq 0} \hat b_{\bf q}\mathrm e^{-i {\bf q}\cdot {\bf r}_j}\nonumber \\
&=& \sqrt{n} + \frac{1}{\sqrt{V}} \hat b_{{\bf q}=0}\nonumber \\
&&\mkern30mu + \frac{1}{\sqrt{V}}\sum_{{\bf q}\neq 0} u_{\bf q} \hat \gamma_{\bf q} \mathrm e^{-i {\bf q}\cdot {\bf r}_j} + v_{-\bf q}^*\hat \gamma^\dagger_{-{\bf q}} \mathrm e^{+i {\bf q}\cdot {\bf r}_j}.\nonumber
\end{eqnarray}

In the TWA~\cite{Blakie:2008iss}, the occupation of momentum mode ${\bf q}$ is described by a set of independent identically distributed complex-valued Gaussian random variables $\gamma_{\bf q}$, analogous to the quantum mechanical operators $\hat{\gamma}_{\bf q}$, such that the semiclassical condensate field is decomposed as
\begin{eqnarray}
\label{eq:TWA_ansatz}
a_j &=& \sqrt{n} + \frac{1}{\sqrt{V}}\sum_{{\bf q}\neq 0} u_{\bf q} \gamma_{\bf q} \mathrm e^{-i {\bf q}\cdot {\bf r}_j} + v_{-\bf q}^*\gamma^*_{-{\bf q}} \mathrm e^{+i {\bf q}\cdot {\bf r}_j}.\nonumber\\
\end{eqnarray}
Since any Gaussian distribution is uniquely determined by its only two non-vanishing moments, we can choose the mean and variance of $\gamma_{\bf q}$ to correctly recover the true quantum mechanical fluctuations within the quadratic Bogoliubov theory~\cite{Blakie:2008iss}.

We can now use the TWA ansatz~\eqref{eq:TWA_ansatz} as an initial condition for the GPE. Notice that this ansatz wavefunction is no longer uniform in space, due to the random contribution from the excited modes, and thus the time evolution needs to be computed in real space. Thus, within the TWA, the uniform condensate wavefunction is seeded with the correct quantum fluctuations, which ensures that the dynamics is captured correctly at the semiclassical level.

To compute expectation values of observables, we produce an ensemble of initial states $a_j$, time-evolve each state separately up to the time of interest, compute the observable for each realisation $a_j(t)$ separately, and perform an ensemble average in the very end. For example, the momentum distribution of the excitations within the TWA is given by
\begin{equation}
n_{{\bf q}\neq 0}(t) = \overline{\quad \left| \frac{1}{\sqrt{V}}\sum_j a_j(t) \mathrm e^{i{\bf q}\cdot {\bf r}_j}\right|^2 \quad },
\end{equation}
where $\overline{(\cdot)}$ stands for averaging over the random variable $\gamma_{\bf q}$.

While the TWA captures the effects of quasiparticle collisions to some extent, it should be noted that this formulation of quantum dynamics is not exact. One of the reasons for this is that it uses Gaussian distributed random variables, whereas in general the true quantum distribution from the BHM has nonzero higher-order moments beyond the Bogoliubov approximation.

\section{\label{sec:numerics} Details of the Numerical Simulations}

\subsection{Homogeneous 2D System}

While it is possible to simulate a 3D \emph{homogeneous} system, the strong harmonic trap along the $z$-axis present in the experiment effectively confines the system to a quasi-2D geometry. We therefore model the experiment by one lattice and one transverse degree of freedom. While the squeezed continuous $z$ direction is irrelevant for the parametric instability effects within BdG theory, effects of the additional $z$-confinement are expected to be felt in the TWA simulations only at later times, beyond the regime of agreement between BdG and TWA (see Fig.~6 in the main text).

For the BdG and WCCA simulations, we use a condensate density of $n=20$ to initialize the system in the Bogoliubov ground state. To determine the condensate wavefunction for the TWA, we use imaginary time evolution to find the lowest-energy state of the GPE in the presence of weak interactions, and consider an ensemble of $10^3$ TWA realizations, performing an additional averaging over the phase of the drive $\varphi$ to reduce the effects of the initial kick. 

For all simulations, the momentum in the $x$ direction (lattice) is discretized according to the quantization condition
$q^x = 2\pi l/(N^x_\mathrm{sites} d)$ with $l\in[1,2,\dots,N^x_\mathrm{sites}]$. To simulate numerically the $y$ (continuous) direction we apply the following auxiliary discretization: $q_\perp = 2\pi l_\perp/ (N^\perp_\mathrm{sites} d_\perp)$ with $l_\perp\in[1,2,\dots,N^\perp_\mathrm{sites}]$, imposing an upper cutoff on the values of perpendicular momentum at $2\pi/d_\perp$. We checked that increasing this cutoff does not change the results, since high-momentum states have high energy and remain unaffected by the dynamics. 

The volume of the finite 2D system system is $V=N^x_{\text{sites}}\,N^\perp_{\text{sites}}$, where $N^x_{\text{sites}}$ and $N^\perp_{\text{sites}}$ are the number of sites along the respective axis. Setting $N^x_\mathrm{sites}=81$ and $N^\perp_\mathrm{sites}=80$ modes, we find that the perpendicular momentum grid is dense enough for the system to be in the thermodynamic limit already at $d_\perp=d=1$. We further make sure the results remain unchanged with increasing the number of modes. 

Effects due to the weak confining harmonic trap are not considered in these simulations, and are addressed separately [cf.~Appendix~\ref{sec:SI_trap}]. We use QuSpin~\cite{Weinberg:2017chs,Weinberg:2018vqs} to simulate the dynamics.

\subsection{\label{sec:numericsImhom}Inhomogeneous 1D system}

The computational complexity of simulations in the presence of harmonic confinement limits the reachable system sizes. Unlike homogeneous systems, the inhomogeneity induced by the trap leads to a break-down of momentum conservation and couples all momentum modes. Hence, to determine the Bogoliubov mode functions, needed for the initial condition for the EOM, one has to exactly diagonalize a matrix of size $N_\mathrm{sites} \times N_\mathrm{sites}$, with $N_\mathrm{sites}$ the total number of space points. It is the size of this matrix, together with the additional bottlenecks for solving the corresponding equations of motion, which set the current limit on the reachable system sizes. For these reasons, simulating the dynamics of a 3D system in the thermodynamic limit ($N_\mathrm{sites}\sim 10^6$) in the presence of a trap and subject to periodic modulation, is not feasible without introducing further approximations. We therefore limit the theoretical discussion of effects due to the harmonic trap to 1D lattices.

To obtain the numerical data in the presence of a harmonic trap potential, we simulate the BdG equations for a 1D lattice of $L_x/d=201$ sites~\cite{NoteMomCons}.  We first perform imaginary time evolution in the presence of the trap to determine the exact initial condensate profile. This inhomogeneous profile results in an inhomogeneous effective interaction strength $g(x)$, which is used to compute the Bogoliubov modes in the absence of the periodic drive used as an intial condition~\cite{Hutchinson1997s}. Finally, the time-periodic BdG dynamics is simulated by solving the real-space variant of Eq.~\eqref{eq:BdGE_EOM} with the additional replacement $g\to g(x)$.

\section{\label{sec:constants}Constants and Model Parameters}

\begin{tabular}{ c | c | c }
	Variable & SI Units & Theory Units \\
	\hline
	$d$  & $425\,$nm & 1 $d$ \\
	$J$  & 108 Hz$\times h$ & 1 $J$ \\
	$E_R$ & $7.1\,$kHz$\times h$ & 65.4 $J$ \\
	$k_L$ & 739 $\cdot$ 10$^4$ $\frac{1}{\mathrm{m}}$ & $\pi \frac{1}{d}$\\
	$\hbar$ & 1.05 $\cdot$ 10$^{-34}$ Js & 1 $\hbar$ \\
	$m$  & 6.5 $\cdot$ 10$^{-26}$ kg & 0.07543 $\frac{\hbar^2}{J d^2}$ \\
	$U_0$ & 2.3 $\cdot$ 10$^{-51}$ Jm$^3$ & 0.4144 $J d^3$ \\
	$\zeta$ & 2.08 & 2.08 \\
	$N$ & 3.7 $\cdot$ 10$^5$ & 3.7 $\times$ 10$^5$ \\
	$\omega_r$ & 2$\pi\times$26 Hz & 0.24 $\frac{J}{\hbar}$ \\
	$\omega_z$ & 2$\pi\times$204 Hz & 1.75 $\frac{J}{\hbar}$ \\
\end{tabular}
\newline 

\vspace{2cm}


\begin{thebibliography}{16}
\bibitem [{Par()}]{PartNum}%
  \BibitemOpen
  \href@noop {} {}\bibinfo {note} {This makes the dynamics sensitive to the
  value of the chemical potential used, which is counterintuitive, given that
  the starting point is the BHM where particle number is
  conserved.}\BibitemShut {Stop}%
\bibitem [{Not()}]{NoteMomCons}%
  \BibitemOpen
  \href@noop {} {}\bibinfo {note} {Note that the momentum-space description is
  not useful here since the trap breaks momentum conservation.}\BibitemShut
  {Stop}%
\end{thebibliography}
\end{document}